\documentclass[12pt,letterpaper]{JHEP3}
\pdfoutput=1
\usepackage{amsmath,amssymb,graphicx} %use drftcite in draftmode
\usepackage{epsf}

\newcommand{\beq}{\begin{eqnarray}}% can be used as {equation} or  {eqnarray}
\newcommand{\eeq}{\end{eqnarray}}

\newcommand{\centeron}[2]{{\setbox0=\hbox{#1}\setbox1=\hbox{#2}\ifdim
\wd1>\wd0\kern.5\wd1\kern-.5\wd0\fi
\copy0

\kern-.5\wd0\kern-.5\wd1\copy1\ifdim\wd0>\wd1
                                       \kern.5\wd0\kern-.5\wd1\fi}}
\newcommand{\ltap}{\>\centeron{\raise.35ex\hbox{$<$}}
                               {\lower.65ex\hbox{$\sim$}}\>}
\newcommand{\gtap}{\>\centeron{\raise.35ex\hbox{$>$}}
                               {\lower.65ex\hbox{$\sim$}}\>}

\newcommand{\ltsimeq}{\raisebox{-0.6ex}{$\,\stackrel
        {\raisebox{-.2ex}{$\textstyle <$}}{\sim}\,$}}
\newcommand{\gtsimeq}{\raisebox{-0.6ex}{$\,\stackrel
        {\raisebox{-.2ex}{$\textstyle >$}}{\sim}\,$}}

\newcommand\ZZ{\hbox{\zfont Z\kern-.4emZ}}
\font\zfont = cmss10 %scaled \magstep1

\newcommand{\ls}{l_s}
\newcommand{\Rads}{R_{\rm AdS}}
\newcommand{\Vol}{{\rm Vol}}
\newcommand{\vol}{{\rm v}}
\newcommand{\LTev}{\Lambda_{\rm TeV}}
\newcommand{\LUV}{\Lambda_{\rm UV}}

\title{
Randall-Sundrum and Strings
}
\author{Matthew Reece\\ 
Princeton Center for Theoretical Science\\
Princeton University, Princeton, NJ 08544, USA\\
E-mail: \email{mreece@princeton.edu}}
\author{Lian-Tao Wang\\
Department of Physics, Princeton University, Princeton, NJ, 08540\\
E-mail: \email{lianwang@princeton.edu}}

\abstract{
We investigate
stringy excitations in  Randall-Sundrum effective theories for electroweak symmetry breaking arising from embedding in string theory. RS is dual to a confining gauge theory, which we expect to have ``QCD strings," or color flux tubes. Stringy constructions of RS-like theories allow us to investigate the mass of these string states, which typically grows with a small fractional power of the number of colors $N$ of the dual gauge theory. There are two known strong constraints on $N$ for RS-like theories. The first arises from demanding that the Standard Model gauge couplings do not have a Landau pole at low scales. The second arises from demanding that the first-order confining phase transition in the early universe is able to proceed without leaving an empty universe, i.e. that the rate of bubble nucleation is not too small. We find that these constraints on $N$ imply that string states are generically 
at most a factor of a few heavier than the lightest KK states, and we cannot self-consistently remain in the limit $N, \lambda \gg 1$. We examine various string constructions of AdS or RS-like backgrounds, including orbifolds, theories on M5-branes,  theories on D4-branes, and the recent F-theory construction of Polchinski and Silverstein. In every case we find that there are strong bounds on the mass of new stringy states.
 We briefly discuss  important phenomenological implications due to the presence of such light stringy excitations, such as precision electroweak and flavor observables, as well as collider signals. 
}

\begin{document}

%\renewcommand{\thefootnote}{(\arabic{footnote})}

%%%%%%%%%%%%%
\section{Introduction}
\label{sec:intro}
\setcounter{equation}{0}
\setcounter{footnote}{0}
%%%%%%%%%%%%%

Supersymmetry and compositeness are two main scenarios of possible solutions of the hierarchy problem. 
The AdS/CFT correspondence \cite{maldacena, GKP, WittenAdS} has opened up new perspectives on the compositeness scenario. Many important model building efforts have been carried out in the Randall-Sundrum scenario \cite{RS}, which is a weakly coupled dual description of the strongly coupled near conformal dynamics which generates a composite scale around TeV \cite{HoloPheno}. A pedagogical introduction, with references to much of the literature, can be found in Ref. \cite{Csaki:2005vy}. Being a low energy effective description, the RS setup needs an UV completion to become a full solution to the hierarchy problem.  In this article, we will focus on implications of string theory as such a UV completion. 

There are in principle two different embeddings of  the Standard Model in such a UV completion. It is possible that the Standard Model gauge symmetries are emergent from strong dynamics. This is similar to RS1 with all SM gauge and matter fields localized on the TeV brane. UV completion of such theories is perhaps similar to the Klebanov-Strassler scenario \cite{KS}. We will not focus on this scenario in this article.

We will concentrate on possibility that the SM interactions come from weakly gauging the global symmetries of the nearly conformal strong dynamics. 
This is a common property of large classes of technicolor and composite Higgs models.  
Due to the AdS/CFT correspondence, this is equivalent to putting the SM gauge fields in the RS bulk \cite{bulkgauge,custodialRS}. Many properties of the TeV composite states can be inferred, such as the existence of KK excitations of the SM particles. In this paper, we would like to argue that a stringy UV completion generically implies the existence of an additional set of stringy resonances which is not significantly heavier than the commonly studied KK resonances.  These may be thought of as the ``QCD strings" or confining flux tubes of the RS sector.

Our arguments consist of two main steps. First, the AdS background is a dual description of some large N gauge theory. The ratio of four-dimensional masses of stringy resonances and KK modes is
\beq
\frac{m_{\rm str}}{m_{\rm KK}} = \frac{\Rads}{\ls}  =  a N^{\delta}
\eeq
where $\Rads$ is the AdS radius, and $\ls = \sqrt{\alpha'} $ is the sting length.\footnote{We explain in detail how bulk length scales relate to four-dimensional masses in Appendix \ref{app:4d5d}.}  Typically, $a$ is of $\mathcal{O}(1)$, and  $\delta$ is some small fractional power. For example, $a = (4 \pi g_s)^{1/4}$ and $\delta=1/4$ in the original AdS$_5 \times S^5$ (in the calculable limit $\Rads \gg \ls$).

Next, we provide two separate arguments that for a consistent and phenomenologically viable model in this category, $N$ is at most $\mathcal{O}$(10). First of all, we expect the Standard Model gauge couplings to remain perturbative above the TeV scale. Otherwise, they will be an integral part of the strong dynamics rather than weakly gauging its global symmetry. Generically, we expect the $\beta$-function for the Standard Model gauge couplings is proportional to $N$. Hence, perturbativity  implies an upper bound on $N$. Second, the RS sector has a deconfinement phase transition at the TeV scale. If $N$ is too large, this transition is too strongly first-order, and cosmology has an empty universe problem. Both of these arguments have been made in the 
%JHEP
Randall-Sundrum context \cite{PomarolRunning, HoloPheno, Contino:2002kc, Agashe:2005vg, CreminelliEWPT,EmptyUniverse}.
The main focus of this paper will be providing details of these two arguments, and investigating possible ranges of the ratio $m_{\rm str}/m_{\rm KK}$. We focus on various stringy implementation of this type of model, setting aside orthogonal issues that make stringy UV completion of nonsupersymmetric RS a challenge \cite{Strassler:2003ht, Kachru:2009kg}. 

Of course, string theory on Ramond-Ramond backgrounds and highly curved spaces is not under theoretical control. Our approach is to perform a self-consistency check. We examine the relationship between $N$ and $m_{\rm str}/m_{\rm KK}$ in the weakly curved (super)gravity limit where all the string excitations are decoupled.  We will see that applying the bounds on $N$ implies strings states are light enough that the gravity limit with only KK modes only has a very limited regime of validity, and string excitations should be relevant for low energy observables. 

Remarks that bounds on $N$ set limits on $m_{\rm str}/m_{\rm KK}$ have appeared in a few places: in the context of hidden valley models, it was pointed out that even quite large $\lambda \sim 100$ is not enough to completely decouple strings \cite{Strassler:2008bv}. More recently, two studies of particular types of stringy resonances in Randall-Sundrum models have appeared. The first gave roughly the above argument and claimed that a spin-3/2 stringy excitation of the top quark is likely to be the lightest string resonance \cite{Hassanain:2009at}, while the second studied spin-2 reggeons \cite{Perelstein:2009qi} that were argued to lie not too far above the TeV scale. 

Instead of string theory, one might also consider a UV completion of RS based on the idea of deconstruction \cite{Deconstruction,DeconstructingRS}. However, as this type of completion does not incorporate gravity, it is not clear whether it is necessary to consider approximating a warped space. Hence, we will treat this direction, while certainly interesting to explore, as orthogonal to the type of scenario considered here.  

This paper is organized as follows. In section~\ref{sec:boundsN}, we review the bounds on $N$ from perturbativity and the first order phase transition. In Section~\ref{sec:scales}, we study specific bounds on the ratio $m_{\rm str}/m_{\rm KK}$ in a set of possible string theory UV completions. We provide an alternative argument based on the behavior of a 4D confining potential in Section~\ref{sec:generic}, which leads to a similar conclusion. We outline the implications to the phenomenology of TeV scale new physics in Section~\ref{sec:toymodel}. Section~\ref{sec:Conclusion} contains our conclusion.

%%%%%%%%%%%%%
\section{Phenomenological Bounds on Number of Degrees of Freedom}
\label{sec:boundsN}
\setcounter{equation}{0}
\setcounter{footnote}{0}
%%%%%%%%%%%%%

In this section we will review some bounds on the number of TeV-scale degrees of freedom that are coupled to the Standard Model in technicolor-like theories. 

\subsection{Bounds from the Perturbativity of SM Gauge Couplings}

We consider a new strongly-interacting  sector which has a group of global symmetries, some of which are gauged by the Standard Model gauge bosons. For instance, a QCD-like technicolor sector would have a global symmetry SU(2)$_L \times$ SU(2)$_R \times$ U(1)$_B$, a subgroup of which is gauged by the SM SU(2)$_L \times$ U(1)$_Y$. We imagine that the SM gauge interactions are a weak perturbation of the dynamics of the strongly-interacting sector. At leading order in the weak SM gauge couplings, the Standard Model gauge beta functions are modified by the two-point function of the global symmetry current in the strongly-interacting sector. For instance, if the strongly-interacting sector is a pure CFT, then we would have (exactly in the CFT, and at leading order in the SM gauge coupling):
\beq
\int~d^4 x~e^{-i q \cdot x} \left<J_\mu(0) J_\nu(x)\right>_{CFT}  =  -\frac{b_{CFT}}{16\pi^2} \left(q^2 g_{\mu \nu} - q_\mu q_\nu\right) \log q^2.
\eeq
Then, the  SM gauge coupling at low energies $Q$ is given by
\beq
\frac{8 \pi^2}{g^2(Q)} =  \frac{8\pi^2}{g^2(\LUV)} +\left(b_{SM} + b_{CFT}\right) \log \frac{\LUV}{Q}
\eeq
If the strongly-interacting sector is not conformal, then in general the running of the gauge coupling induced by the strong sector is not precisely logarithmic. This subtlety is not important for our purpose, since above a threshold set by the scale of the mass gap in a confining strong sector, the running is logarithmic to good approximation.

Now the key argument is that $b_{CFT}$ should not be too large, or the Standard Model gauge interactions will very rapidly become strongly interacting above the scale of the lightest states in the strong sector. 
We wish to avoid hitting a Landau pole at scales that are just above the weak scale, and possibly even far above the weak scale.  For one, it would imply that the SM gauge interactions are not really a weak perturbation of the strong sector, so that there is a complicated theory of multiple strongly-interacting gauge groups with dynamics that we cannot solve. 
For another, it would suggest that the SM gauge bosons should probably be thought of as composites, as in Seiberg duality \cite{Seiberg:1994pq}. 
This is a perfectly reasonable possibility to consider \cite{SMasComposite,KS,SMcascade}, but it would lead us to theories that are conceptually very different. %LTW 
%from technicolor. 
%JHEP
The existence of a Landau pole bound is implicit in the early RS literature that understood the matching of 4D and 5D gauge couplings as corresponding to logarithmic running due to CFT degrees of freedom \cite{PomarolRunning, HoloPheno}, and was made fully explicit in discussions of GUTs in RS \cite{Contino:2002kc, Agashe:2005vg}.
Now, the precise bound on $b_{CFT}$ depends on which scale a Landau pole is acceptable at, as shown in Figure \ref{fig:bCFTbound}. For example, if the mass scale where the strongly interacting sector begins is at 1 TeV, and we wish to explain a large hierarchy, up to say 10$^{15}$ GeV, then to avoid a Landau pole for SU(2)$_L$ we have:
\beq
b_{CFT} \leq \frac{8 \pi^2}{g^2(\LTev)} \frac{1}{\log 10^{12}} + \frac{10}{3} \approx 10.
\label{eq:bcftbound}
\eeq
The $10/3$ is from $b_2 = -10/3$ in the Standard Model, after subtracting the Higgs contribution of $+1/6$, since its role is replaced 
%LTW
by the strongly-interacting sector. If some of the Standard Model fermions are composites, or if there is still an elementary Higgs field, the computation changes appropriately, but the bound remains ${\cal O}(10)$. The bound weakens in the ``little Randall-Sundrum" scenario that gives up on explaining the large hierarchy, with flavor bounds in mind \cite{LittleRS}. 

\FIGURE[!ht]{
\includegraphics{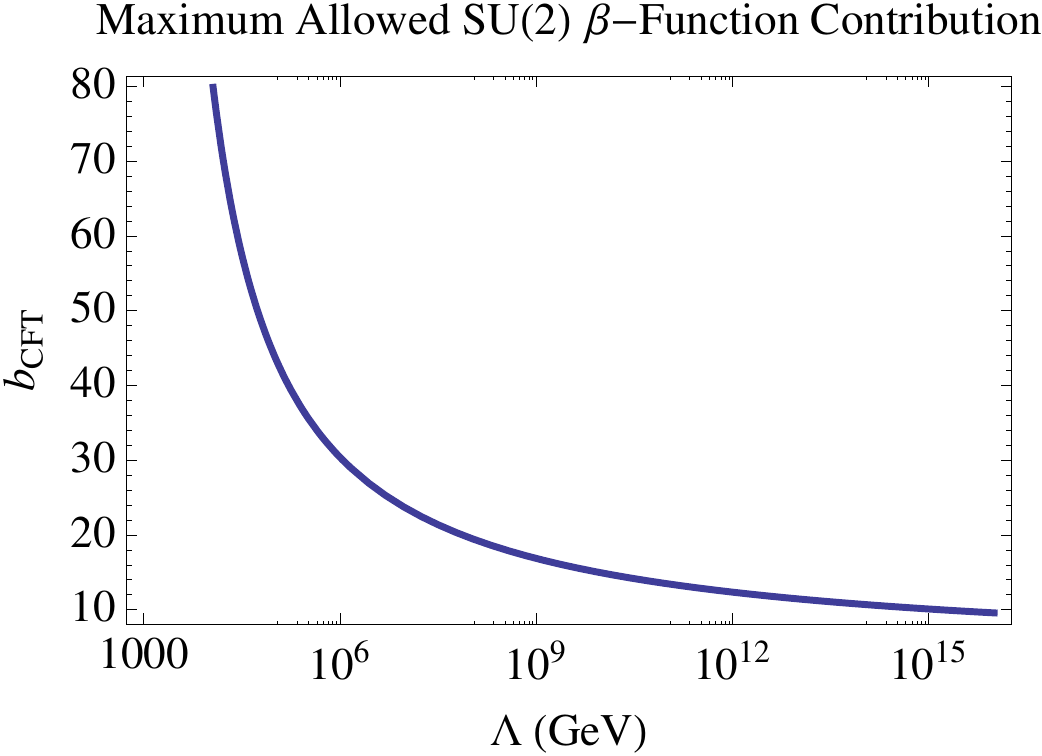}
\caption{Bound on $b_{CFT}$ as a function of the scale $\Lambda$ below which we forbid a Landau pole.}
\label{fig:bCFTbound}}

%JHEP
For a theory of electroweak symmetry breaking,  we need not assume that the SU(3) color interaction gauges a global symmetry of the %LTW
 strongly-interacting sector, so the most generic constraint is the one from SU(2)$_L$. On the other hand, most Randall-Sundrum models in the literature {\em do} assume that the full Standard Model gauge group is a subgroup of the global symmetry group of %LTW
 the strongly-interacting sector, because it gives a nice way of dealing with flavor. 
%JHEP
In these models, there is a constraint on $b_{CFT}$ for the SU(3)$_c$ currents as well, which numerically is very similar (the QCD coupling is larger but its SM beta function is more asymptotically free).

Naively, if the strongly-interacting sector is some large-$N$ gauge theory, we expect that $b_{CFT}$ arises from matter that is charged under the SU($N$) symmetry and under a global symmetry. If that matter is in the fundamental of SU($N$), then $b_{CFT} \sim N$, and in other representations $b_{CFT}$ scales even faster with $N$. In this case,  the bound we have discussed (in the case of a large hierarchy) implies that $N \ltsimeq 10$. However, there are cases where $b_{CFT}$ is ${\cal O}(1)$ rather than ${\cal O}(N)$  or larger \cite{GaiottoMaldacena}. We will revisit this point below, but first we turn to an independent phenomenological bound which applies to the {\em total} number of degrees of freedom.

\subsection{Bounds from First-Order Phase Transition}
\label{sec:firstorder}

A second bound on the number of new degrees of freedom near the TeV scale arises from cosmology. It depends on the assumption that the universe was at some point at a temperature above the scale of the deconfinement transition in the new strongly-interacting sector. The highest temperature for which we have convincing data is around 4 MeV from nucleosynthesis \cite{TReheat}, so if one assumes that the universe was never reheated to scales above the deconfinement transition, one can avoid this bound. On the other hand, unlike the bound on $b_{CFT}$, the cosmological bound gives a constraint on the {\em total} number of degrees of freedom in the strong sector, not just those that are gauged by the SM interactions. In particular, it gives a bound on $N$ even in cases where $b_{CFT} \sim 1$.

A clear and general discussion of the electroweak phase transition, and associated cosmological difficulties, for large $N$ gauge theories was given in Ref. \cite{EmptyUniverse} (see also earlier remarks on cosmology and finite temperature in RS in Ref. \cite{HoloPheno}). The key point is that for large $N$ gauge theories, the confining phase transition is first-order (where, in practice, ``large $N$" means $N \gtsimeq 3$), and so proceeds by bubble nucleation. In the deconfined phase, the free energy scales with $N^2$ (the number of gluons). The phase transition becomes more strongly first order with larger $N$, and the transition rate goes as $e^{-N^2}$, so if $N$ is large the bubbles never collide. Such theories have the ``empty universe problem."

To understand the scaling, we should review the physics of the phase transition in the dual gravity picture. It is a Hawking-Page transition \cite{HawkingPage}, which was related to deconfinement transitions in gauge theory early in the AdS/CFT literature \cite{WittenAdS,WittenBH}. In the canonical ensemble for AdS spaces, there are two solutions with the right asymptotics: a ``thermal AdS" (i.e. AdS compactified on a thermal circle), and an AdS-Schwarzschild solution. In AdS spaces dual to confining gauge theories, where the space ends at $z = z_{\rm IR}$, the phase transition occurs when the black hole horizon of the AdS-Schwarzschild solution falls behind the wall associated with confinement. Thus, at high temperatures, we have a gauge theory plasma, dual to a thermal AdS space; at low temperatures, we transition to the theory of hadrons with masses set by the location of the hard wall. This geometric picture makes it clear that the critical temperature should be associated with the scale of the light ``KK mode" hadrons, not with parametrically heavier stringy states. In particular, for the hard wall model, the critical temperature can be calculated (by evaluating regulated 5D actions $I_{5D}$ for the two classical geometries, proportional to free energies) to be \cite{Herzog:2006ra}:
\beq
T_c  = 2^{1/4}/(\pi z_{\rm IR}).
\eeq
At the phase transition, the entropy density changes from ${\cal O}(N^2)$ at high temperatures to ${\cal O}(1)$ at low temperatures, reflecting the binding of gluons and quarks into hadrons.\footnote{M5-brane theories have ${\cal O}(N^3)$ degrees of freedom, but for the moment our discussion assumes a more traditional gauge theory.} The change in vacuum energy scales as ${\cal O}(N^2) z_{\rm IR}^{-4}$. In particular, for the hard-wall model, one can calculate the change in vacuum energy at the transition using $E_{\rm vac} = -\frac{\partial}{\partial \beta} \log Z = \frac{\partial}{\partial \beta} I_{5D}$ (where $Z$ is the partition function and $\beta$ the radius of the thermal circle) \cite{BallonBayona:2007vp}:
\beq
\Delta E_{\rm vac} = \frac{16 M_5^3 \Rads^3}{z_{\rm IR}^4} = \frac{8}{\pi^2} c \frac{1}{z_{\rm IR}^4}.
\eeq  
Here $c$ is the central charge of the CFT dual to our AdS background. We have a bubble nucleation rate $\Gamma \sim z_{\rm IR}^{-4} e^{-{\cal O}(N^2)}$ and a Hubble scale set by $H^2 M_{\rm Pl}^2 = \Delta V \sim N^2 z_{\rm IR}^{-4}$. Unfortunately, we can't analytically calculate the bounce action for the instanton that creates bubbles of the confined phase within the deconfined plasma, so we can at best give a bound up to an order-one number. The bound set by requiring that bubbles collide is $\Gamma \gtsimeq H^4$, i.e.
\beq
a_0 z_{\rm IR}^{-4} \exp^{-a_1 c} > c^2 z_{\rm IR}^{-8} M_{\rm Pl}^{-4},
\eeq
with $a_0, a_1$ unknown order-one numbers. That is, as a bound on the central charge:
\beq 
N_{\rm d.o.f.} = c \ltsimeq \frac{1}{a_1} \left(4 \log(M_{\rm Pl} z_{\rm IR}) + \log a_0 - 2 \log c \right).
\eeq
For convenience, we will simply quote this as a bound at $a_0 = a_1 = 1$ and $M_{\rm Pl} z_{\rm IR} = 10^{16}$:
\beq
N_{\rm d.o.f}=c \ltsimeq 140,
\label{eq:cbound}
\eeq
with the understanding that this is subject to uncertainties and unknown order-one dependence on details of the background. 
%JHEP
(In particular, ``order-one" in this context means not scaling with $N$; without an understanding of the bounce action, we are not able to make definite statements about large factors like 4$\pi$, so the cautious reader interested in concrete numbers may prefer to focus on the Landau pole bound.)
For typical string backgrounds, $b_{CFT} \sim N$ and $c \sim N^2$, so the two bounds \ref{eq:bcftbound} and \ref{eq:cbound} are comparably strong.

Our discussion has focused on the case of hard-wall models, and we have not asked how the geometry is stabilized. The electroweak phase transition in Randall-Sundrum theories stabilized by the Goldberger-Wise mechanism was considered in detail in refs. \cite{CreminelliEWPT, RandallServant}, and \cite{Nardini:2007me}. Such theories have a radion parametrically lighter than other modes, and its effective potential can be analyzed, leading to some surprises. In particular, the scaling of various quantities with $N$ is not always as expected from simple field theory considerations. We expect that the hard wall estimate above is a very good guide to theories where a mass scale is introduced explicitly through relevant operators. For theories with logarithmic running leading to confinement, e.g. Klebanov-Strassler \cite{KS}, we also expect the estimates above to be a better guide than GW-like models. An approximate calculation has been carried out for a Klebanov-Tseytlin throat, which bears out this expectation \cite{Hassanain:2007js}. In particular, the bound thus obtained is: $N_{\rm IR}^2 = \frac{27 \pi^2}{4 g_{s}^2} \left(M_5 \Rads(z_{\rm IR})\right)^3 \leq 21$, where $N_{\rm IR}$ should be interpreted as the number of effective degrees of freedom at the bottom of the throat. This suggests that our bound \ref{eq:cbound} may be an overestimate. 

%%%%%%%%%%%%%
\section{Scales in AdS in string compactifications}
\label{sec:scales}
\setcounter{equation}{0}
\setcounter{footnote}{0}
%%%%%%%%%%%%%
In this section, we study a set of string compactifications with warped throats. In each case, we calculate the ratio $m_{\rm str} / m_{\rm KK}$ as a function of $N$. To incorporate the bounds from perturbativity, we also compute the low energy gauge coupling as a function of string coupling and $N$. 

We emphasize that our main result is that $m_{\rm str}$ and $m_{\rm KK}$ are generically not {\em parametrically} separated if $N$ is not formally large. Although we exhibit numerical estimates of $m_{\rm str}/m_{\rm KK}$ for simple cases, we will demonstrate our point by showing its dependence on the scaling of the underlying parameters in more involved examples. 

\subsection{Generalities}
\label{sec:generalities}

We begin by considering a general theory that looks like AdS$_5$ times a $d$-dimensional internal space, where $d$ might not be 5. The 5d Planck scale is:
\beq
M_5^3 & \propto & \frac{\Vol_{d}}{g_s^2 \ls^{3+d}}
\label{eq:planck5d}
\eeq
At the same time, calculation of the conformal anomaly using gravity in AdS$_5$ leads to \cite{Henningson:1998gx}:
\beq 
\label{eq:dof}
N_{\rm d.o.f} = c = 2 \pi^2 M_5^3 \Rads^3
\label{eq:centralcharge}
\eeq
From equations \ref{eq:planck5d} and \ref{eq:dof}, we obtain 
\beq
\label{eq:ratio_general}
\frac{\Rads}{\ls} \propto \left( \frac{g_s^2 N_{\rm d.o.f}}{\vol_d} \right)^{\frac{1}{d+3}}
\eeq
where $\vol_d = \Vol_d / \Rads^d$ is the volume of the internal manifold in the units of $\Rads$.
This suggests that the way to achieve an AdS curvature radius that is large in string units is to try to shrink the internal volume (relative to the AdS size) or to increase the number of degrees of freedom. In fact, the volume of the internal space and the number of degrees of freedom are not decoupled, as we will see momentarily. We also see the potential difficulty of achieving this due to the fractional power $1/(d+3)$ on the right hand side.

\subsection{Einstein Manifold}

We begin with the simplest class of examples with AdS$_5 \times M_5$ where $M_5$ is an Einstein manifold, which includes the original case $M_5 = S^5$.  AdS$_5$ is the near horizon limit of a stack of $N$ D3 branes. There are $N$ units of flux on the internal manifold $M_5$.  The Standard Model matter and gauge fields can be added by including D7 branes wrapping around a 3-dimensional submanifold of $M_5$, for example an equatorial $S^3 \subset S^5$ \cite{KarchKatz}. 
%JHEP
(For a discussion of RS-like phenomenology with D7 branes, see ref. \cite{Gherghetta:2006yq}.)

Reducing the 10d gravity action on $M_5$, we obtain 5d Planck constant as 
\beq
M_5^3 = \frac{1}{(2 \pi)^7 g_s^2 \ls^8} \Vol_{M_5}.
\eeq
Consider first the simplest case where $M_5 = S^5$ with radius $R$. 
In this case, we have the well-known relation 
\beq
\label{eq:s5}
\Rads^4=R^4 = 4 \pi g_s N \ls^4. 
\eeq
We can also obtain the same relation by considering the number of degrees of freedom and using AdS/CFT. We expect $N_{\rm d.o.f} \propto N^2$ on the CFT side. Using Eq.~\ref{eq:dof}, we have
\beq
N^2 \propto \Rads^3 M_5^3 \propto \frac{\Rads^8}{g_s^2 \ls^8},
\eeq
which again implies Eq.~\ref{eq:s5}. 
This relation highlights the condition of having a tractable effective field theory: string states are parametrically heavier than the supergravity states, by a factor $(g_s N)^{1/4} \gg 1$. 

For a more general Einstein manifold $M_5$, stabilized by $N$ units of flux (arising from D3 branes at the tip of a cone over $M_5$) we have \cite{Gubser:1998vd}
\beq
\label{eq:m5}
\Rads^4 = 4 \pi N g_s \ls^4 \frac{\pi^3}{\vol_{M_5}},
\eeq
where $\vol_{M_5}$ is the volume of $M_5$ in units of $\Rads$, and $\vol_{S^5} = \pi^3$. Therefore, the mass ratio of string modes and KK modes in this background is
\beq
\label{eq:ratio_m5}
\frac{m_{\rm str}}{m_{\rm KK}} = \frac{\Rads }{\ls}  = \left( 4 \pi g_s N \frac{\pi^3}{\vol_{M_5}} \right)^{1/4}. 
\eeq
As discussed earlier, this ratio can be increased either by increasing $N$  or $\pi^3/\vol_{M_5}$, or both. Typical examples of $M_5$ are:  $S^5$ with $\vol_{S^5} = \pi^3$; $S^5/Z_2$ with $\vol_{S^5/Z_2} = \pi^3/2$; or the conifold $T^{1,1}$ with $\vol_{T^{1,1}} = 16 \pi^3 /27$, all of which contribute at most an order 1 factor to this ratio. S-duality of IIB string theory implies that $g_s = 1$ is the largest sensible value of the coupling, so we cannot increase $m_{\rm str}/m_{\rm KK}$ indefinitely by increasing $g_s$.

For the purpose of setting the cosmological bound, it is more convenient to use a different, equivalent formulation of equation \ref{eq:ratio_m5}, making use of eq. \ref{eq:centralcharge}:
\beq
c = \left(\frac{\Rads^4}{8\pi l_s^4 g_s}\right)^2 \left(\frac{\vol_{M_5}}{\pi^3}\right).
\eeq
Note that this is a general result which can also be used in other kinds of compactifications. It depends only on geometry, not on a stabilization mechanism. Then the bound (\ref{eq:cbound}) from avoiding the empty universe problem gives a constraint:
\beq
\frac{m_{\rm str}}{m_{\rm KK}} \ltsimeq \left(140 \times 64 \pi^2 \frac{\pi^3}{\vol_{M_5}}\right)^{1/8} = 4.2 \left( \frac{\pi^3}{\vol_{M_5}}\right)^{1/8}.
\label{eq:cboundeinstein}
\eeq
If we had chosen $c_{max}$ in the range 10 to 1000, the coefficient 4.2 could vary from 3.0 to 5.3. Thus the only way to achieve a large hierarchy given the empty universe bound is to make the internal space small in units of the AdS radius.

Next we consider the bound from the $\beta$-function constraints from perturbativity of SM gauge couplings beyond the TeV scale. We will demonstrate this explicitly in this example.  We start with the DBI action of the D7 brane 
\beq
S_{\rm DBI} = - \tau_7 \int d^8 \sigma~{\rm tr} \sqrt{-{\rm det} (G_{\alpha \beta} + 2 \pi \alpha' F_{\alpha \beta})}
\eeq
where $\tau_7 = 1/(g_s (2\pi)^7 \ls^8)$, which leads to kinetic term\footnote{Note that because we consider a nonabelian gauge group (namely SU(2)$_L$), there is a factor of $1/2$ arising from the trace over gauge indices that is not present in the abelian case.}  
\beq
\mathcal{L}_{\rm kin} = -\frac{1}{2 g_7^2} \int {\rm tr} F^2, \ \ g_7^2 =  2g_s (2 \pi)^5 \ls^4.
\eeq
After integrating over the 3D submanifold $M_3$ which the D7 brane wraps around, we obtain the 5-D gauge coupling
\beq
\frac{\Rads }{g_5^2} = \Rads \times \frac{{\rm Vol}_{M_3}}{g_7^2}   = \frac{\vol_{M_3}}{2\pi^2} \frac{2\pi^2}{2g_s (2 \pi)^5}\left(\frac{\Rads}{\ls} \right)^4,
\eeq
where $\vol_{M_3}$ is the volume of $M_3$ in units of $\Rads^3$. This implies that 
\beq
b_{CFT} = 8\pi^2\frac{\Rads}{g_5^2} = \frac{\vol_{M_3}}{2\pi^2} \left(\frac{\Rads}{\ls}\right)^4 \frac{1}{4\pi g_s}.
\label{eq:bcftgen}
\eeq
This is a general result depending only on geometry, which can also be used in other compactifications. Now, using eq. \ref{eq:m5} to specialize to the case of $N$ D3-branes on a cone over an Einstein manifold, we find:
\beq
b_{CFT} = \frac{\vol_{M_3}}{2\pi^2} \frac{\pi^3}{\vol_{M_5}} N.
\eeq
This is consistent with expectations that $b_{\rm CFT} \sim N$, since the matter charged under the global symmetry is a bifundamental of the global symmetry group and the SU($N$) gauge theory dual to our AdS space. In particular, for the Karch-Katz example of D7 branes wrapping an $S^3 \subset S^5$, we find $b_{CFT} = N$, as expected for flavor hypermultiplets. 

Using Eq.~\ref{eq:bcftgen}, we find:
\beq
\label{eq:ratio_beta}
\frac{m_{\rm str}}{m_{\rm KK}} \ltsimeq \left(4\pi g_s \frac{2 \pi^2}{\vol_{M_3} } \left(\frac{8 \pi^2}{g^2 (\LTev)} \frac{1}{\log (\LUV / \LTev)} + \frac{10}{3}\right)\right)^{1/4} \ltsimeq 3.3 \left( g_s  \frac{2 \pi^2}{\vol_{M_3} }\right)^{1/4},
\eeq
where we have used Eq. \ref{eq:bcftbound} for numerical concreteness in the last step. Notice that this bound depends only on the volume of the cycle wrapped by the D7-brane, and not on $\vol_{M_5}$; indeed, we did not use eq. \ref{eq:m5} at all.

One possible way of getting a large ratio of $m_{\rm str}/m_{\rm KK}$ is having $\vol_{M_3} $ to be parametrically smaller than the volume of an $S^3$ with radius $\Rads$. 

\subsection{The Klebanov-Strassler Cascading Geometry}

We have been discussing AdS spaces, imagining that we simply truncate them as in RS to obtain a ``hard wall" model of confinement (which may be a particularly good approximation for theories in which confinement is driven by a relevant operator \cite{PS1}). Spaces that solve the hierarchy problem, however, will tend to have geometries that are cut off in a more gentle way, since we expect that they involve marginal or nearly-marginal operators. The canonical example of a string construction of such a theory is the Klebanov-Strassler geometry \cite{KS, DualityCascadeLectures}, in which the number of degrees of freedom runs logarithmically with energy scale until the tip of the throat. The topology of the internal dimensions of KS is $S^2 \times S^3$, with the $S^2$ shrinking to zero size at the end of the throat. One might wonder if, because the geometry near the end of the throat resembles a compactification on $S^3$, the scaling of various quantities will be very different from the AdS$_5 \times X^5$ examples we have discussed.

The K-S geometry is dual to an ${\cal N}=1$ SU($N+M$)$\times$SU($N$) gauge theory with bifundamentals $A_{1,2}$, antibifundamentals $B_{1,2}$, and a superpotential $\lambda \det_{r,u} (A_r B_u)$ preserving an SU(2)$\times$SU(2)$\times$U(1) global symmetry. This theory exhibits a sequence of Seiberg dualities $N \to N-M$ which reduce the number of degrees of freedom. Correspondingly on the gravity side there is a running AdS radius:
\beq
R^4(r) = \frac{81}{8} (g_s M)^2 \alpha'^2 \log(r/r_s).
\eeq
Here $r_s$ is the coordinate at the tip of the throat. There are $M$ units of 3-form flux on the $S^3$, which is constant throughout the geometry, while the amount of 5-form flux and of $B$-field flux on the $S^2$ scale as $\log(r/r_s)$. Relative to the theory on AdS$_5 \times S^5$, the main change in scaling relations is that $N$ has been replaced with a quantity $\sim g_s M^2 \log(r/r_s)$. Here $M$ is related to the number of degrees of freedom in the deep IR, where the theory becomes simply an SU($M$) gauge group that confines and generates a mass gap. We can think of $g_s M^2 \log(r/r_s)$ as a slowly-running $N(r)$ analogous to the value of $N$ for AdS$_5 \times S^5$. It determines the (slowly running) central charge, as well as the mass of string states relative to the AdS curvature radius, in the same manner as $N$. The cosmological argument regarding phase transitions bounds the value of $N(r)$ near the tip of the throat. A more detailed calculation, as we noted earlier, was performed in Ref. \cite{Hassanain:2007js} for the Klebanov-Tseytlin throat.

The Landau pole argument, on the other hand, reflects an integrated number of degrees of freedom over some region in the throat. Flavor symmetries can be obtained by adding D7 branes in the Klebanov-Strassler throat, and solutions have been found for backreacting smeared flavor branes \cite{D7FlavorInThroat}. For our purposes, it is enough to consider probe branes, and to simply use an approximate metric:
\beq
ds^2 = \left(\frac{R(z)^2}{z^2}\right)\left(\eta_{\mu \nu} dx^\mu dx^\nu + dz^2\right) + R(z)^2 d\Omega^2,
\eeq
with $d\Omega^2$ the metric on $T^{1,1}$. Because the D7 branes wrap three internal dimensions, we will have an important factor of $R(z)^3$ in the beta function calculation:
\beq
\int d^4x dz\sqrt{g} {\rm Vol}(\Sigma) g^{\mu\rho}g^{\nu\sigma}F_{\mu\nu}F_{\rho\sigma} \sim  \int d^4x dz\left(\frac{R(z)}{z}\right) R(z)^3 F_{\mu\nu}F_{\rho\sigma},
\eeq
which is precisely what is needed to find the expected power $R(z)^4 \sim N(z)$ scaling with the running number of degrees of freedom. Thus, unlike RS, for which we observe logarithmic running of the beta function, in the KS case we have $\log^2$-running, i.e., $b_{CFT}$ itself grows logarithmically with energy scale. Thus, up to order-one factors, in the KS case all arguments are exactly as they were in AdS$_5 \times S^5$, but with $N$ corrected to a running $N(r)$. The ratio of the mass of 4d string and KK states is determined by $R(r)/l_s$ at the bottom of the throat, so the cosmological bound translates essentially exactly as in AdS$_5 \times S^5$. The Landau pole bound can be thought of as bounding $b_{CFT}$ in the middle of the throat (i.e. at $\sqrt{\LTev \LUV}$).

\subsection{Orbifolds and $Y_{p,q}$}

So far we have considered spaces in which all internal dimensions are of roughly the same size as the AdS radius.
Since spaces in which $\vol_{M_5}$ is much smaller than $\Rads^5$ may lead to larger values of $m_{\rm str}/m_{\rm KK}$ , we now investigate the cases in which some internal dimensions are taken to be much smaller than others. The first such examples are orbifolds, which inherit their properties in a direct way from AdS$_5 \times S^5$ \cite{orbifold}. We can view the 5-sphere as a circle fibered over ${\mathbb C \mathbb P}^2$. A freely acting discrete group ${\mathbb Z}_k$ rotates the circle fiber, so we can take a quotient and obtain a smooth space AdS$_5 \times S^5/{\mathbb Z}_k$. The internal space still has radius $\Rads$ in four directions, but along the fiber direction its radius has been reduced to $\Rads/k$. The corresponding dual is an $SU(n)^k$ quiver gauge theory (with $N = nk$). Because the AdS directions are unchanged by the orbifold operation, the ratio of 4d masses $m_{\rm str}/m_{\rm KK}$ still obeys Eq.~\ref{eq:s5}. 
  However, the internal volume is smaller, so both the central charge and $b_{CFT}$ for a D7-brane wrapped on the small cycle are smaller, and we have loosened the bound at no cost!
  However,  for the construction to make sense in string theory, the identified points cannot lie closer together than the string scale. Hence, $k < N^{1/4}$.   Our bounds on $N$ were of order 10, so we can shrink the number of degrees of freedom by a factor $k \sim 2$. 
Since  an even smaller power of this will enter the ratio $m_{\rm str}/m_{\rm KK}$, orbifolding can only change this ratio by a very limited amount.

Another important constraint for realistic models constructed on this background  can be obtained  by considering the nonperturbative instability in  this theory \cite{HOP}. This instability is similar to the original ``bubble of nothing" instability of the Kaluza-Klein vacuum, arising from solutions in which the circle fiber shrinks to zero size at the edge of a bubble \cite{Witten:1981gj}. Despite this instability, similar orbifolds have been discussed as a source of stable non-supersymmetric RS-like spaces in string theory, because if such a throat is embedded in a Calabi-Yau, the lifetime of the space can be longer than the age of our universe \cite{Kachru:2009kg}. However, the decay
width goes as $\Gamma \sim k^9 \LUV^4 \exp\left(-\frac{N^2}{k^8}\right)$, so that if we approach the limit $k \sim N^{1/4}$, the entire space would very rapidly decay. Thus in this case,  a stronger condition $k \ll N^{1/4}$ is not merely necessary for a controlled theory; it is necessary for the very existence of the theory. 

Next, we consider another example in which one internal dimension is much smaller than the others. These are the $Y_{p,q}$ spaces, where $p$ and $q$ are integer labels \cite{Ypq}. The topology of the internal space in this family of solutions is $S^2 \times S^3$, with metric depending on $p$ and $q$. Like the Klebanov-Witten geometry, the $Y_{p,q}$ spaces can be used as the basis for a cascading geometry analogous to Klebanov-Strassler \cite{Herzog:2004tr}. The volume of the $S^2 \times S^3$ in a $Y_{p,q}$ is given by:
\beq
\vol_{Y_{p,q}} = \frac{\pi^3 q^2 \left(2 p + \sqrt{4 p^2 - 3 q^2}\right)}{3p^2 \left(3 q^2 - 2 p^2 + p \sqrt{4 p^2 - 3 q^2}\right)} = \frac{16 \pi^3}{27 p} - \frac{2 \pi^3 q^2}{27 p^3} + {\cal O}(p^{-4}),
\eeq
where in the last step we have taken the limit $p \gg q$. This limit produces a small-volume, anisotropic internal space, and by taking it we might hope to push string states toward heavier scales. However, it turns out that the $Y_{p,q}$ spaces have a geometry that is, for our purposes, similar to that of the orbifolds. In particular, the five-dimensional geometry can be viewed as a circle bundle over $S^2 \times S^2$, and in the limit $p \gg q$ when the overall volume is $\sim 1/p$, the circle fiber direction has length of order $1/p$ and the other directions have length of order one (in units of $\Rads$). Then, just as in the orbifold example, requiring that the circle is larger than the string scale implies that  $p < (g_s N)^{1/4}$, providing a limit to how much one can shrink the geometry while consistently using a IIB string description.

\subsection{Small internal dimensions through tuning}
\label{sec:polchinskisilverstein}

Recently Polchinski and Silverstein have given F-theoretic constructions of AdS spaces with fairly small internal dimensions \cite{PolchinskiSilverstein}. F-theory is essentially just type IIB string theory with D7 branes; the use of D7 branes to construct small internal dimensions is motivated by the observation that they can help to cancel curvature terms. D7 branes lead to gauge fields in the bulk of AdS, which correspond to global symmetries in the dual theory. 
To have a theory of electroweak symmetry breaking, we need  some of the global symmetries of the strongly-interacting sector to be weakly gauged by the Standard Model electroweak gauge group. Therefore, D7 branes could be natural ingredients if one is seeking a phenomenological model. Thus, the Polchinski-Silverstein constructions look like a promising starting point for the pursuit of Randall-Sundrum-like models in string theory. We proceed to make a simple estimate of the ratio $m_{\rm str}/m_{\rm KK}$ in this setup. 

The D7 branes allow one to tune a curvature term in the potential that normally would be of order $1/R^2$, for a radius $R$ of some internal directions, to be of size only $\epsilon/R^2$, with $\epsilon \ll 1$. In the particular AdS$_5$ construction of Ref. \cite{PolchinskiSilverstein}, there are three separate length scales:
\beq
R_f \sim \epsilon \Rads, \ R \sim \epsilon^{1/2} \Rads, \ \Rads^4 \sim \frac{N}{\epsilon^3} \ls^4,
\eeq 
where $R_f$ and $R$ are, respectively, the sizes of a special fibered direction and the remaining 4 directions of the internal manifold. The last relation is parametrically the same as Eq.~\ref{eq:m5}. Much as in the simpler orbifold case, there is a limit to how small we can shrink the internal dimensions. Requiring that the smallest length scale in the construction is above the string length, we find that $\epsilon \gtsimeq 1/N$. 

For the cosmological bound, we use $\vol_{M_5}/\vol_{S^5} \sim \epsilon^3$. If we set $\epsilon \sim 1/N$, then we have $c \sim N^2 \epsilon^{-3} \sim N^5 \ltsimeq c_{\rm max} = 140$ (from Eq. \ref{eq:cbound}) and $\Rads/\ls \sim N$. Thus we have:
\beq
\frac{m_{\rm str}}{m_{\rm KK}} \sim N \sim c_{\rm max}^{1/5} = 2.7.
\eeq
This is to be contrasted with the scaling $\sim c_{\rm max}^{1/8}=1.8$ in the AdS$_5 \times S^5$ case. We see that only a very mild tuning is allowed by the constraints. While the scaling of $c$ is somewhat milder, it does not lead to a significant enhancement of the mass hierarchy.

Next we consider the Landau pole bound. We have $\vol_{M_3}/(2 \pi^2) \sim \epsilon^n$ where $n=3/2$ if the 3-dimensional submanifold the D7 brane wraps around has $\Vol_{M_3} \propto R^3$, and $n=2$ if $M_3$ includes the dimension with size  $R_f$. (The D7-branes that are intrinsically present in the Polchinski-Silverstein F-theory construction wrap the fiber and so have $n=2$. The case $n=3/2$ only makes sense if one can consider probe branes with small backreaction; since $N$ is bounded, this seems unlikely to be realizable.) Then we have:
\beq
b_{CFT} \sim \epsilon^n \frac{N}{\epsilon^3}
\eeq
Taking $\epsilon$ as small as possible, i.e. $\epsilon \sim N^{-1}$, we then have $b_{CFT} \sim N^{4-n}$ and $\Rads \sim N \ls$. Thus our bound is:
\beq
\frac{m_{\rm str}}{m_{\rm KK}} \leq \left(\frac{8 \pi^2}{g^2 (m_Z^2)} \frac{1}{\log (\Lambda_{\rm UV}^2/ \LTev^2)} + \frac{10}{3}\right)^{\frac{1}{4-n}},
\eeq
where the exponent $1/(4-n)$ is $1/2$ if $M_3$ wraps the circle fiber (as we expect in the F-theory construction) and $2/5$ if not, to be contrasted with $1/4$ in the simplest example of AdS$_5 \times S^5$. Again, the enhancement of the mass hierarchy is not significant.

\subsection{Scales in D4-brane Theories}
\label{sec:D4}

We see from Eq.~\ref{eq:ratio_general} that we can change the fractional power in the ratio $m_{\rm str} / m_{\rm KK}$ by changing the number of internal dimensions. One family of examples includes theories on D4 branes. Such theories, compactified on a circle with SUSY-breaking boundary conditions, give IR dynamics that is thought to be in the same universality class as pure Yang-Mills \cite{WittenBH}. D8 flavor branes can be added in the bulk to give rise to quarks, giving a QCD-like theory \cite{SakaiSugimoto,SakaiSugimoto2}. In these theories, the relation is $\Rads^3 = \pi g_s N \ls^3$. The difference from Eq.~\ref{eq:s5} results from the fact that  we are only compactifying down to 6D (before we compactify on the final circle, which lives in the {\em boundary} theory and so is of qualitatively different character). 

The evaluation of $\beta$-function here is somewhat more involved since the dilaton is not constant along the warped direction in these theories. The full details have been worked out in Refs.~\cite{SakaiSugimoto,SakaiSugimoto2}. We will just quote relevant results here. The DBI action of a D8 brane in this backrgound is 
\beq
S_{\rm D8} = - \tau'_8 \int d^4 x \int^{z_{\rm UV}}_{- z_{\rm UV}} d z \left(\frac{\Rads}{4 U_z} F_{\mu \nu} F^{\mu \nu} + ...  \right) + \mathcal{O} (F^3),
\eeq
with 
\beq
U_z & \equiv  & U_{\rm KK} \left( 1+ \frac{z^2}{U_{\rm KK}^2 } \right)^{1/3} \\ \tau'_8 & = & \frac{1}{54 \pi^3} M_{\rm KK} N \frac{1}{\ls^2}.
\eeq
$U_{\rm KK}$ parameterizes the size of the AdS throat. $R_{S^1} = M_{\rm KK}^{-1}$ is the size of the circle that the D4 branes wrap around. We have $U_{\rm KK} \sim M_{\rm KK}^{-1} \sim \Rads$ with a choice $M_{KK}^2 \ls^2 = \frac{9}{2} (g_{YM}^2 N)^{-1}$ \cite{SakaiSugimoto2}. The weakly coupled SM gauge bosons correspond to modes with $A_{\mu}(x, z \rightarrow z_{\rm UV} ) \rightarrow 1$.  Performing the $z$ integral, we obtain an estimate for the 4D effective gauge coupling
\beq
\frac{1}{g_4^2} \sim \frac{1}{9 \pi^2} \tau'_8 \Rads^3 \left( \frac{z_{\rm UV}}{U_{\rm KK}}\right)^{1/3}. 
\eeq
Therefore, using $N = (\pi g_s)^{-1}(\Rads / \ls)^3 $, we estimate
\beq
\frac{m_{\rm str}}{m_{\rm KK}} \sim \left[ 54 \pi^4 \frac{g_s}{ g_4^2} \left( \frac{U_{\rm KK}}{z_{\rm UV}}\right)^{1/3} \right]^{1/5}.
\eeq
Although the ``techniquarks" of this theory live on the 3+1 dimensional intersection of D4 and D8 branes, they couple to 4+1 dimensional gluons, and cause the running of the gauge coupling to be power-law rather than logarithmic above the KK scale. While the interpretation changes, and it may not make sense to discuss the value of a 4d gauge coupling in the UV, it is clear that the running is much stronger in this theory and the bound on $b_{CFT}$ is not ameliorated. Models of compositeness built on such theories would be, in a loose sense, RS $\times$ UED. (The TeV extra dimensions are not quite UED, as only a subset of the modes propagate in them; in particular, the SM gauge boson KK modes live on the D8 branes which are localized on the circle.) Such theories have been discussed from a phenomenological viewpoint in Refs. \cite{Carone:2007md, Hirayama:2007hz, Carone:2008rx, Mintakevich:2009wz}.

The Sakai-Sugimoto model can be generalized by placing the D8 branes at a separation $L \ll R_{S^1}$ on the circle the D4 branes are compactified on \cite{Antonyan:2006vw}. The confinement scale will always be of order $M_{KK}$, but in this case the D8 and $\overline{\rm D8}$ branes will meet at $U \gg U_{KK}$, so chiral symmetry breaking occurs at an energy scale well above the confinement scale. Because the chiral symmetry breaking effect is what sets the $W$ and $Z$ masses for electroweak symmetry breaking, this would be a theory with a light glue sector that is only probed by electroweak-scale physics. In other words, it gives an example of a ``hidden valley" \cite{HiddenValley}. (In some ways such a  scenario could resemble the physics of quirks \cite{Quirks} and of the pure-glue hidden valley \cite{GlueHiddenValley}, although the glue here is five-dimensional and involves additional modes associated with the $S^4$.)

The finite-temperature behavior of the Sakai-Sugimoto background is not so different from that of RS as discussed in Section \ref{sec:firstorder}, with a black-hole solution that dominates for $T > \frac{1}{2\pi R_{S^1}}$ in a first-order deconfinement transition \cite{SakaiSugimotoFiniteT}. Thus, regardless of which limit we are in, the cosmology suggests a bound on the total number of degrees of freedom. In the quirk-like limit, with the deconfinement transition occurring at a scale below the electroweak phase transition, the constraint becomes even more difficult to avoid.

\subsection{M5-brane Theories}
\label{sec:M5}

Recently it has been observed that theories built on M5 branes may avoid the Landau-pole constraint \cite{GaiottoMaldacena}. The reason is simply that in M theory the brane action does not carry the factor of $1/g_s$ that gives the scaling with $N$ of the beta-function contribution from a flavor D-brane in the constructions we have been discussing. In particular, it was suggested that a global symmetry could arise from M5 branes wrapping AdS$_5 \times S^1$, with coupling of order $\Rads/R_{S^1}$. A stabilization with the circle radius of order the AdS radius would give only an order-one contribution to the beta function. The known theories along these lines are ${\cal N}=2$ 4d SCFTs that can be realized by compactifying 6d (0,2) SCFTs on Riemann surfaces \cite{Gaiotto:2009we}, so constructing a Klebanov-Strassler-like modification is an open problem.

However, assuming such a construction is possible, a phenomenological application of such theories would still run into the first-order phase transition argument of section \ref{sec:firstorder}. In fact, for such M-theoretic constructions, the constraint is likely to be stronger, because the total number of degrees of freedom scales as $N^3$, not as $N^2$. Thus, it appears that even $N \sim 10$ would run into a sharp phenomenological difficulty, albeit one that might be avoided with a low reheating scale.

In an M5-brane theory, the nonperturbative objects are M2 and M5 branes, and there are no strings, so the question of the mass scale of stringy resonances is somewhat complicated. However, if the geometry contains a circle on which we can wrap an M2 brane, which is extended in another direction, one can view this as a stringy state. (If the theory exhibits confinement, we expect that some sort of flux-tube-like state is present; this seems to be the most natural way to find such a state.) The M2-brane tension scales as $M_{Pl}^3$, so the string tension of an M2-brane wrapped on a circle of radius $\Rads$ is $\Rads M_{Pl}^3 \approx N^{1/3} M_{Pl}^2$, using the relationship %LTW some notations
$\Rads \sim N^{1/3} M^{-1}_{Pl}$. (This arises when all internal dimensions are parametrically the same size, of order $\Rads$.) Thus the string mass scale is $N^{1/6} M_{Pl}$, to be compared to the inverse curvature radius $\Rads^{-1} \sim N^{-1/3} M_{Pl}$. From this we see that the string / KK mode hierarchy in such a model could be of order $\Rads/l_s \sim N^{1/2}$. In particular, Gaiotto and Maldacena discuss a class of theories on genus $g$ surfaces with $c = \frac{N^3}{3} (g - 1)$. Taking $g = 2$ and applying the bound (\ref{eq:cbound}), we find that $\Rads/\l_s \sim N^{1/2} \ltsimeq 2.7$. While a detailed exploration of whether interesting confining backgrounds arise in M5-brane theories would no doubt be interesting, these crude estimates suggest that avoiding the empty universe problem would constrain such theories just as much as those arising from string theory.

\subsection{Weak gravity and hypothetical examples}

The weak gravity conjecture gives a lower bound on the volume of the internal manifold \cite{weakgravity}:
\beq
\Vol_{d}  \gtsimeq g_s \Rads \ls^{d-1}.
\label{eq:weakgrav}
\eeq
It shows as a proof of principle that it is nontrivial to avoid having large internal dimensions. We know no examples that saturate it. In fact, most examples satisfy it with an additional large $N$ factor. Typically, one term in the potential that stabilizes internal dimensions is a flux factor of order $g_s^2 N_{flux}^2/\Vol_d^2$ 
($N_{flux}=N$), measured in string units. Without fine-tuning, all terms in the potential are typically of the same order at the minimum, where the value of the potential must be the cosmological constant $-20/\Rads^2$. Setting these terms to be of the same order, we obtain $\Vol_d \sim g_s N_{flux} \Rads \ls^{d-1}$, safely above the bound by a factor of $N_{flux}$ which must be large for calculability.

Because the volume of the internal manifold is directly related to the central charge, we can ask what a hypothetical example saturating (\ref{eq:weakgrav}) would imply for our bounds. In such an example, we would have $\vol_{M_5} = \Vol_5 \Rads^{-5} \sim g_s \left(\ls/\Rads\right)^4$, and thus $c_{sat} \sim (\Rads/\ls)^4$ is of order the square root of its value in a theory with all internal dimensions of size $\Rads$. Using the cosmology bound conservatively, this could allow $\frac{m_{\rm str}}{m_{\rm KK}} \ltsimeq \left(16 \pi^2 \times 140\right)^{1/4} \approx 12$. The $b_{CFT}$ bound is similarly weakened. If we take $\frac{\vol_{M_3}}{2\pi^2} \sim \left(\frac{\ls}{\Rads}\right)^3$ (i.e. wrap three string-scale internal dimensions), then the bound from equation \ref{eq:ratio_beta} is $\frac{m_{\rm str}}{m_{\rm KK}} \ltsimeq 4 \pi \times 10 \approx 120$. If we take $M_3$ to wrap two dimensions of radius $\ls$ and one of radius $\Rads$, the bound is $\frac{m_{\rm str}}{m_{\rm KK}} \ltsimeq \sqrt{40 \pi} \approx 11$. None of these numbers should be taken very seriously, as we have no construction of a theory that saturates the weak gravity bound. Unlike in earlier examples, we are now dealing with large enough powers of $c$ and $b_{CFT}$ that factors of 2 and $\pi$ are very important, and where these factors go is pure guesswork. Optimistically, one might view this as a challenge for string model-builders: perhaps an example exists which saturates the weak gravity bound, is at low enough curvature to be under calculational control, and has factors of $\pi$ in opportune places to allow a hierarchy $m_{\rm str} \gg m_{\rm KK}$. We know of no general counterargument, but any such example would involve a more delicate stabilization mechanism than simply turning on $N$ units of 5-form flux.

Although we have listed a general internal dimension $d$ in eq. \ref{eq:weakgrav} and section \ref{sec:generalities}, we expect that only critical string theories (or M-theory) will give weakly curved backgrounds. We could consider a noncritical string, but in this case, the defect in central charge of the worldsheet theory generically sources string-scale curvature. In such theories one tends to have no separation at all between $\Rads$ and $\ls$. Nonetheless, they still obey $\Vol_d \sim g_s N_{flux} \Rads \ls^{d-1}$, because $g_s N_{flux} \sim {\cal O}(1)$ (see, e.g., \cite{KlebanovMaldacena}).

%%%%%%%%%%%%%%%%%%%%%%%%%%%%%%%%%%%%%%%%%%%%%%%%%%%%%%
\section{Generic Large-$\lambda$ Theories And The String Scale}
\label{sec:generic}
\setcounter{equation}{0}
\setcounter{footnote}{0}
%%%%%%%%%%%%%%%%%%%%%%%%%%%%%%%%%%%%%%%%%%%%%%%%%%%%%%

Suppose we have a Randall-Sundrum like theory, but we don't assume that it comes from any known string theory. Do we then lose the argument that %LTW
$\Rads ^4 \sim g_s N l_s^4$? In fact, we still have a route to such an argument. First, we will assume that RS is dual to some sort of large 't Hooft-coupling, confining gauge theory. Now, because the gauge theory confines, it necessarily contains some sort of string-like flux tubes. (This has been argued fairly persuasively over the last 30 years \cite{confinement}.) So it makes sense to ask how heavy these stringy states are. We will use the static quark-antiquark potential as a proxy for the mass of stringy states.

At short distances, this potential should have some sort of Coulombic behavior. It has been shown that resumming ladder diagrams of one-gluon exchange, and extrapolating the resulting Bethe-Salpeter equation to large $\lambda$, leads to a $-\sqrt{\lambda}/r$ form of the potential \cite{resum}. (We review this argument in Appendix \ref{app:resum}.) Extending such a resummation to a generic confining gauge theory is tricky, but we don't need to do so to extract an interesting result. Motivated by the geometric picture of RS, we assume that short-distance conformal (or nearly conformal) behavior should hold for distances below the IR wall, $r \sim z_{\rm IR}$, so that  $V_{\rm static}(r) \sim -\frac{\sqrt{\lambda}}{r}$ is approximately reliable.  Then all we need to draw a conclusion about the string scale is the concavity of the static potential, as proven rigorously by Bachas \cite{Bachas}. At very large $r$, we expect that, due to confinement, $V_{\rm static}(r) \sim \sigma r$, where $\sigma$ is a string tension. (We assume the large-$N$ limit, so screening is not relevant.) Then concavity of the potential tells us that:
\beq
\sigma \ltsimeq \frac{\sqrt{\lambda}}{z_{\rm IR}^2}.
\eeq Now, noting that the mass of a KK mode is $m_{\rm KK} \sim z_{\rm IR}^{-1}$, whereas the mass of a string mode is $m_{\rm str} \sim \sqrt{\sigma}$, this tells us exactly that $m_{\rm str} \ltsimeq \lambda^{1/4} m_{\rm KK}$.

Thus using only concavity of the static potential, resummation of perturbative diagrams, and approximate conformality below the scale $m_{\rm KK}$ of light excitations, we conclude that the asymptotic string tension is bounded above by $\lambda^{1/4} m_{\rm KK}$. This matches the known scaling behavior of ${\cal N} = 4$ SYM, which appears from quite different geometrical considerations (note that the $S^5$ and its volume and associated fluxes played no role in the present discussion, for which we did not even assume the existence of a string theoretic dual!). Resumming perturbative diagrams is, of course, not a rigorous way to obtain correct information about large-$\lambda$ theories. But it is reassuring that without using input from string theory, we are able to obtain the same qualitative conclusion. It suggests that the results from known string theories could be robust when extrapolated to other, as yet unknown, strongly coupled gauge theories.

%%%%%%%%%%%%%
\section{Phenomenological implications}
\label{sec:toymodel}
\setcounter{equation}{0}
\setcounter{footnote}{0}
%%%%%%%%%%%%%

The presence of a low string scale will alter the phenomenology of Randall-Sundrum models in a number of ways. Most obviously, string resonances could be directly produced at colliders \cite{Hassanain:2009at, Perelstein:2009qi}. Higher-spin particles with the quantum numbers of various Standard Model states would be expected, as well as higher-spin closed string states that do not carry Standard Model quantum numbers (but which might e.g. decay to $W$ and $Z$ bosons). There are also indirect effects of string states on low-energy precision observables. We turn to this issue now.

\subsection{Precision electroweak}

As in any technicolor or composite Higgs theory, an important constraint on Randall-Sundrum models arises from the Peskin-Takeuchi precision electroweak observables, especially the $S$ and $T$ parameters \cite{techniEW}, which precision data prefer to be near zero.  In RS, these parameters are calculable, 
and in the RS-type model with SM matter fields on the IR brane (from the dual point of view, they are composite) both $S$ and $T$ are large and negative \cite{CsakiErlichTerning}. The $T$ parameter problem is easily resolved by requiring that the bulk gauge group is large enough to contain a custodial symmetry; it is often taken to be SU(2)$_L \times$ SU(2)$_R \times$ U(1)$_{B-L}$ \cite{custodialRS}. Because the $T$ parameter is protected by a symmetry, we expect that provided a stringy completion of RS continues to respect custodial symmetry (i.e. as long as SU(2)$_R$ is a good gauge symmetry in the bulk), the $T$ parameter is safe.

The $S$ parameter is more dangerous because it is not protected by any symmetry (apart from SU(2)$_L \times $ U(1)$_Y$ itself). It receives contributions both from the symmetry-breaking dynamics of the strongly-interacting sector and from effects related to the couplings of the SM fermions to gauge fields \cite{Disguising, Cacciapaglia:2006pk}. The result that $S < 0$ with SM fermions on the IR brane is driven by the compositeness of the SM fermions. If fermions are localized on the UV brane, i.e. thought of as external spectators to the CFT dynamics, then the symmetry-breaking dynamics of %LTW
the strongly-interacting sector dominates the contributions to the $S$ parameter and it is positive in all known calculable models \cite{HiggslessOblique, positiveS}. The only known way to achieve Higgsless RS theories with an $S$ parameter consistent with the current bounds is ``fermion delocalization," which requires tuning the bulk mass of the fermion to achieve a profile that is nearly orthogonal to the gauge boson KK modes \cite{delocalized}.

There is a sum rule relating the contribution to $S$ from symmetry-breaking %LTW
dynamics to the spectrum of vector and axial vector mesons:
\beq
S = 4 \pi \sum_{n} \left(\frac{f_{V,n}^2}{m_{V,n}^2} - \frac{f_{A,n}^2}{m_{A,n}^2}\right),
\label{eq:Ssum}
\eeq 
where $m_{V(A),n}$ is the mass of the $n$th vector (axial vector) meson and $f_{V(A),n}$ is the corresponding decay constant. In RS, the decay constants are computed as %LTW some changes below R->Rads in several places
$f_n^2 m_n^2 = \frac{\Rads }{g_5^2} \left(\lim_{z\to 0} \frac{ \partial_z \phi_n(z)}{z}\right)^2$, where $\phi_n(z)$ is the normalized wavefunction of the $n$th KK mode. (In non-RS backgrounds, a similar statement is true, but the $\frac{1}{z}$ factor may be modified and a factor of the volume of internal dimensions may be present.) The $\Rads /g_5^2$ scaling shows that $S$ will be proportional to $b_{CFT}$, i.e. the contribution of the technicolor sector to the running of SU(2) gauge coupling. The remaining factors give order-one coefficients that, in the absence of a complete string theory in which we understand all orders in $\alpha'$, are not calculable. The asymptotic Euclidean behavior  $\Pi(Q^2) \sim \log Q^2$ implies a scaling $\lim_{n \to \infty} f^2_n / m^2_n \sim 1/n$ for the resonances, so progressively higher terms in the sum rule will contribute less than the first few resonances.
 We expect that the masses and decay constants of the lightest KK resonances will change only a small amount due to mixing with string states. The contribution from the string resonances should be similar to that of the higher KK modes. However, we do not expect that such contributions are negligible. 
 For instance, in a simple RS Higgsless model, $S = 6\pi \frac{\Rads }{g_5^2}$ \cite{HiggslessOblique}.
  The lightest vector and axial KK vector in eq. \ref{eq:Ssum} contribute $\approx 3.5\pi \frac{\Rads}{g_5^2}$, while the second pair contributes a further $0.9\pi \frac{\Rads}{g_5^2}$. Thus, although the lightest states contribute most of $S$,
the total $S$ could change by an order-one factor by altering the spectrum of higher excitations.

What does this imply for the consistency of RS-like models with precision constraints? 
 Satisfying precision electroweak constraints in Higgsless RS models already required a tuning. Positive $S$ from electroweak symmetry breaking was canceled by negative contributions from the fermion couplings. Stringy physics will alter the calculations in ways that we can't compute, but the upshot of this should be that a different choice of fermion bulk mass will be ``ideally delocalized" and lead to small $S$. Such a change in the preferred bulk mass may even be desirable, from the dual point of view. In the dual, elementary fermions mix with composite fermionic operators with the same quantum numbers, as in the Kaplan mechanism \cite{Kaplan:1991dc}, and tuning $S \to 0$ requires that the composite operators have particular dimensions. 
 %JHEP
 The original Higgsless models prefer a nearly-flat bulk profile, which corresponds to a composite fermionic operator $\Psi$ either of dimension 3/2, or of dimension 5/2 and coupling to an elementary fermion \cite{AdSFermions}. The former is at the unitarity bound and not a plausible dimension for an operator in a strongly-coupled gauge theory, unless anomaly-matching forced the theory to generate massless bound states. The latter is potentially viable, with its interpretation as a small or large anomalous dimension depending on the operator content of the underlying theory (for instance, for an operator of the form $\rm{tr}(\phi\psi)$, it would coincide with the engineering dimension, which would be a surprise in a strongly-coupled CFT). An alternative approach to fermion masses would be to avoid composite operators with the same quantum numbers and instead attempt to holographically realize extended technicolor. This approach was pursued in the D4-D8 construction of Ref. \cite{Hirayama:2007hz}, with SM fermions living at the intersection of the D8 branes and another set of flavor D4 branes. The location of the flavor branes appears to play a similar role to delocalization, allowing a tuning of fermion couplings to cancel the $S$ parameter. To summarize, stringy corrections will induce order-one corrections and alter the tunings needed to achieve consistency with experiment, but will not change the viability of the calculable effective theories.

In theories that are not in the Higgsless limit, but have a composite Higgs (pseudo-Goldstone or not), the remarks we make here about tuning fermion masses to cancel $S$ are less relevant, because the contributions of the sum rule \ref{eq:Ssum} to $S$ are $v^2/m_{KK}^2$ suppressed. Such models can have a more interesting flavor structure because the fermion masses are not tightly constrained by cancellation of $S$.\footnote{Flavor in RS is an active field; see \cite{RSflavor} for an incomplete list of work that might serve as a starting point for the literature.} 
Similar remarks apply: flavor violation generically occurs in RS theories through couplings to KK modes. Any of the various proposals for suppressing couplings of elementary fermions to these composite states to protect flavor should also be expected to suppress couplings to the stringy states. The string states will give quantitative, but not qualitative, corrections to conclusions about the viability of the theory in the face of precision flavor data.

\subsection{Probing Stringy states at colliders}

One of the most interesting possibilities to consider is the direct production of stringy states at colliders. 
Studies of probing TeV scale fundamental string states in other scenarios at colliders have been conducted \cite{string_collider}.  QCD-like string states in the RS scenario provide a new possibility with distinct phenomenology. 

A schematic drawing of the spectrum of some of the lower lying states is shown in Figure~\ref{fig:spectrumschematic}. As we have already emphasized, the precise spectrum of string states in this background is not known. Therefore, although we expect they are not much heavier than the KK modes, we can not predict exactly their mass scales.  We do expect there are higher spin modes with spin going up with mass, similar to the Regge behavior
($J \sim m^2$) of the QCD resonances.

Depending on the structure of the theory, stringy excitations of different particles might or might not exist. Bulk modes like gravitons are universally present, and there will corresponding higher-spin excited closed string states. These are the ``glueballs" of the strong sector and will be most relevant at colliders if SM states are composite; for instance, if the top quark is largely composite, such states may be produced in association with tops 
\footnote{The study of this signature based on KK gravitons has been carried out \cite{glueball}. }. Any technicolor-like theory will have a set of resonances associated with the SU(2)$_L \times $U(1)$_Y$ global symmetries (or enlarged SU(2)$_R$ global symmetry), which can be thought of as stringy resonances of the $W$ and $Z$ bosons and the photon. The lighest new states might be expected to be spin-2 $W$, $Z$, and photons \cite{Perelstein:2009qi}, and one might see processes like a spin-2 $W^\pm$ decaying to $W^\pm + \gamma$ or a photon plus a Kaluza-Klein $W$. Most excitations will be out of reach of the LHC, but given a sufficiently powerful collider, a new spectroscopy would open up, reviving the era of partial-wave analyses, Dalitz plots, and other tools familiar from the time when new QCD resonances were discovered frequently. 

\FIGURE[!ht]{
\includegraphics[scale=0.5]{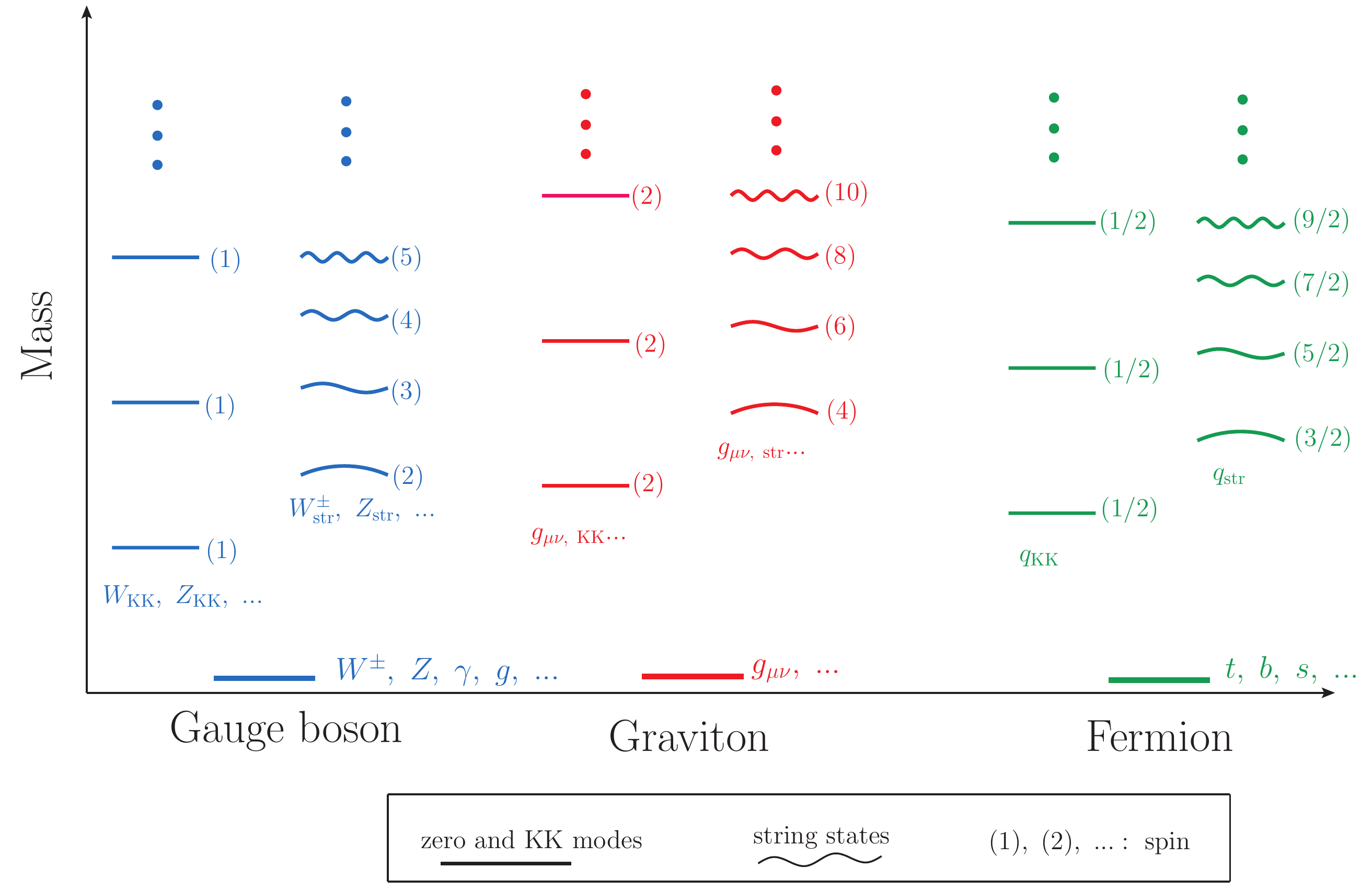}
\caption{
A schematic drawing of the spectrum of lower lying states in the RS scenario. Both the KK modes and string excitations are shown. We have also displayed some of the possible spins of KK resonances and of the string states. Although we have only shown a tower of KK and string resonances on top of each Standard Model state, there can of course be multiple towers with somewhat different masses. Whether there should be KK and string excitations for the Standard Model fermions is model dependent. Note that since the exact spectrum of the string states in this background is not known, both the absolute scale of the string states and splitting between them can only be interpreted as schematic. }
\label{fig:spectrumschematic}}

Other features of the spectrum are more dependent on model-building details. If fermion masses are addressed by the Kaplan mechanism, so that the strong sector has operators with the quantum numbers of Standard Model fermions, these will also have stringy excitations, and in particular the spin-$3/2$ top quark may be an interesting signature \cite{Hassanain:2009at}. On the other hand, an extended technicolor approach to fermion masses would lack such excitations, since the SM fermions would then be purely elementary particles. Kaluza-Klein modes on internal dimensions will generically exist, but are only guaranteed for the closed string states. If open-string states like the $W$ and $Z$ live on flavor branes that wrap internal cycles, then they will have KK modes on those cycles as well. Such states would complicate the spectroscopy, but the distinguishing feature of a string state versus a KK state is that high spin is available.

Optimistically, one could hope that discovering a few resonances at the LHC would be incentive enough to build an even higher energy collider.
At yet higher energies, where the resonances begin to smear out into a continuum, the interesting question is whether partonic behavior of the strong sector will be visible.
 The most important distinguishing feature of large 't Hooft coupling physics, as opposed to small 't Hooft coupling physics, is that partons are never visible. One way to think about this is that partonic branching is so strong, due to the large coupling, that one can never probe an individual parton without it splitting into a large number of soft, small-$x$ partons. Hadrons at large $\lambda$ are bags of soft muck \cite{PolchinskiStrasslerDIS, PolchinskiStrassler}. If Randall-Sundrum constructions could be made phenomenologically viable with a large separation between KK and string scales, this would be the physics at a post-LHC collider. But because, as we have seen, the string states are relatively light, one should expect that partonic physics still applies. This might mean that the physics at high energies is QCD-like, with asymptotic freedom or at least asymptotic conformal behavior with small $\lambda$. But the bounds are not quite so strong, and perhaps the most interesting outcome (from the perspective of learning about strong dynamics) would be a nearly-conformal theory with $\lambda$ of order 1 (or order 4$\pi$) at higher energies: not so small that jetty, QCD-like physics applies, and not so large that spherical events and dual AdS geometry  (as in Ref. \cite{Strassler:2008bv, ads_qcd}) apply. All we know about physics in this regime is speculative \cite{Strassler:2008bv}.

%%%%%%%%%%%%%
\section{Discussion}
\label{sec:Conclusion}
\setcounter{equation}{0}
\setcounter{footnote}{0}
%%%%%%%%%%%%%

In this paper, we have studied the string states in theories of electroweak symmetry breaking in the Randall-Sundrum scenario.  
In particular, we have discussed constraints on the string scale arising from embedding in string theory. By investigating a set of string theory compactifications with warped throats, and imposing phenomenological constraints, we argued that the string states cannot be much heavier than the KK modes. The presence of such light string states, in addition to the often considered KK resonances, can alter the low energy phenomenology significantly, including new signals at colliders and new contributions to electroweak and flavor obervables. 
Although we have focused on RS with SM gauge fields in the bulk, the constraints we have considered can apply to more general RS constructions of strongly-coupled physics. RS hidden sectors could have similar cosmological constraints, for example.

Any real string construction of RS would also face a number of challenges we haven't discussed, including understanding stability of non-supersymmetric backgrounds and avoiding singlet relevant operators that would prevent the theory from solving the hierarchy problem \cite{Strassler:2003ht, Kachru:2009kg}. Our bounds suggest that these problems must be solved in a regime where $\alpha'$ corrections are important, if they are to yield phenomenologically viable solutions. An interesting challenge would be to try to construct a background that saturates the weak gravity bound, having the smallest possible internal dimensions.

%%%%%%%%%%%%%%%%%%%%%%%%%%
\section*{Acknowledgments}

We would like to thank Nima Arkani-Hamed, Roberto Contino, Csaba Cs\'aki, Juan Maldacena, Joe Polchinski, Alex Pomarol, and Matt Strassler for useful discussions and comments. M.R. thanks Juan Maldacena for emphasizing the possibility that M5-brane theories could evade arguments based on the Landau pole. 
%JHEP
We thank Kaustubh Agashe for useful and insightful comments on the preprint. M.R. thanks the Galileo Galilei Institute for Theoretical Physics in Florence, Italy for its hospitality while a portion of this work was completed, and the PCTS for its support. L.-T.W. is supported by the National Science Foundation under grant PHY-0756966 and the Department of Energy (D.O.E.) under Outstanding Junior Investigator award DE-FG02-90ER40542.

\appendix
\section{Relating 5D and 4D Masses}
\label{app:4d5d}

We have claimed that the ratio of masses of four-dimensional string and KK mode states is related to the ratio of the bulk AdS curvature radius and string length:
\beq
\frac{m_{str}}{m_{KK}} = \frac{\Rads}{l_s}.
\label{eq:4d5drel}
\eeq
In this appendix we review the basic facts about holography that lead to this relation. In the bulk, masses of states, measured in units of the AdS curvature radius, are related to the dimensions of operators in the boundary theory. For instance, in AdS$_5$, a scalar dual to an operator of dimension $\Delta$ has mass-squared $m^2 \Rads^2 = \Delta(\Delta - 4)$. Bulk string states correspond to operators that have very large dimension in the boundary.

To compute masses of 4d resonances, we work with the Randall-Sundrum background, although the result will be general. We take the metric to be AdS$_5$ truncated at an IR wall:
\beq
ds^2 = \left(\frac{\Rads}{z}^2\right)\left(\eta_{\mu\nu} dx^\mu dx^\nu + dz^2\right),~~~~0 \leq z \leq z_{IR}.
\eeq
(We may take a UV boundary condition $z_{UV} \leq z$, but for describing just a strongly-coupled %LTW
sector not coupled to elementary fields or gravity, we can send $z_{UV} \to 0$.) For a scalar field of bulk mass $m_5$, we solve for 4D modes with the ansatz $\phi(x,z) = e^{i q \cdot x} \varphi(z)$:
\beq
\partial_z \left(\left(\frac{\Rads}{z}\right)^3 \partial_z \varphi(z)\right) + q^2 \left(\frac{\Rads}{z}\right)^3 \varphi(z) - m_5^2 \left(\frac{\Rads}{z}\right)^5 \varphi(z) = 0.
\eeq
The normalizable solutions (at $z \to 0$) are $\varphi(z) = c_0 z^2 J_\nu(q z)$ with $\nu = \sqrt{4 + m_5^2 \Rads^2} = |\Delta - 2|$. The masses of modes will be determined by a boundary condition at $z = z_{IR}$; for convenience, let us take a Dirichlet boundary condition $\varphi(z_{IR}) = 0$.

For the states we refer to as ``KK modes," $\Delta \sim {\cal O}(1)$ and the bulk mass $m_5 \sim {\cal O}(\Rads^{-1})$. (If we were working with a gauge boson rather than a scalar, for instance, we could have a conserved current with $\Delta = 3$ and $m_5 = 0$, whereas a scalar of dimension 3 has $m_5^2 = -3 \Rads^{-2}$.) When $\Delta \sim {\cal O}(1)$, the masses of the light KK modes are of order $z_{IR}^{-1}$; for instance, if we take $\nu = 1$, then the first zero is at $m_{4d} = 3.83 z_{IR}^{-1}$.

For string states, however, we have very massive bulk fields, $m_5 \Rads \sim \frac{\Rads}{l_s} \gg 1$. In this case, $\nu \gg 1$ and the smallest root of the Bessel function is \cite{WatsonBessel}:
\beq
m_{4d} \approx \left(\nu + 1.856 \nu^{1/3} + {\cal O}(1)\right) z_{IR}^{-1}.
\eeq
Thus we see that for large values of $m_5 \Rads$, or equivalently large operator dimensions $\Delta$, the mass of the lightest 4d state created by the operator is of the order $\Delta\times z_{IR}^{-1}$. In particular, the ratio of 4d masses of the lightest modes created by two operators is of the same order as the ratio of bulk masses of the fields corresponding to those operators. This establishes equation \ref{eq:4d5drel}.

It isn't obvious that solving the two-derivative action for a very heavy field should be a good approximation to masses of excited string states. However, one can find the same result by imagining the behavior of a long, semiclassical excited string state in the bulk. Such a string will fall to the ``bottom" of the AdS geometry and hit the wall at $z = z_{IR}$, where it will correspond to a 4d state with mass given by the warped-down string scale at the wall. This scaling also reproduces equation \ref{eq:4d5drel}, and the consistency of the two viewpoints gives us confidence that this scaling is completely generic.

\section{Resummation}
\label{app:resum}

We briefly review the resummation argument of Ref. \cite{resum}, the applicability of which to non-susy, QCD-like theories was recenty discussed in Ref \cite{Pak:2009em}. This argument computes an approximate strong-coupling potential $V_{static}(r)$ in the short-distance, conformal regime. We choose Feynman gauge, with a gluon propagator $D_{\mu\nu}(x) = \eta_{\mu \nu}\frac{\alpha_s}{x^2}$. We begin with a Wilson loop giving the static potential between two quarks separated by a distance $L$; specifically, we consider a trapezoidal Wilson loop with (Euclidean time) edges of length $T_1$ and $T_2$, both much greater than $L$. Call this Wilson loop $W(T_1, T_2; L)$. The potential we are interested in is given by $V_{static}(L) = -\lim_{T \to \infty} \frac{1}{T} \log W(T, T; L)$. Resumming one-gluon exchanges between the two temporal legs gives a simple Dyson equation for the integral:
\beq
W(T_1, T_2; L) = 1 + \int^{T_1}_0 dt_1 \int^{T_2}_0 dt_2 W(t_1, t_2; L) \frac{\lambda}{4\pi^2 \left(L^2 + (t_1 - t_2)^2\right)},
\eeq
with boundary conditions $W(0, T; L) = W(T, 0; L) = 1$. This integral equation can be converted into a differential equation $\partial_{T_1} \partial_{T_2} W(T_1, T_2; L) =  \frac{\lambda}{4\pi^2 \left(L^2 + (T_1 - T_2)^2\right)} W(T_1, T_2; L)$. This equation is separable and one can find that
\beq
V_{static}(r) = -\frac{\Omega_0}{r},
\eeq
where $\Omega_0$ is the ground-state solution to a Schr\"odinger equation
\beq
\left(-\frac{d^2}{dx^2} - \frac{\lambda}{4\pi^2 (x^2 + 1)} \right) \psi(x) = -\frac{\Omega_0^2}{4} \psi(x).
\eeq
The full derivation and further discussion may be found in Refs. \cite{resum, Pak:2009em}. When $\lambda$ is large, the potential term is large over a range of $x < \sqrt{\lambda}$, and one may approximate $\Omega_0$ by solving the equation expanded around $x = 0$, so the solution resembles a simple harmonic oscillator with $\Omega_0 = \frac{\sqrt{\lambda}}{\pi}$. In the opposite limit, $\lambda \ll 1$, the potential term is small for any $x$, and can be modeled as a $\delta$-function spike at $x = 0$, giving $\Omega_0 = \frac{\lambda}{4\pi}$.

This was the remarkable result of Ref. \cite{resum}, that summing up ladder diagrams gives the correct qualitative behavior, with a potential interpolating between $V_{static}(r) \sim -\frac{\lambda}{r}$ at small coupling and $\sim -\frac{\sqrt{\lambda}}{r}$ at strong coupling. (For circular Wilson loops, Ref. \cite{DrukkerGross} argued that such a resummation gives an {\em exact} answer for ${\cal N}=4$ SYM.)


\begin{thebibliography}{99}

\bibitem{maldacena}
J.~M.~Maldacena,
  %``The large N limit of superconformal field theories and supergravity,''
  Adv.\ Theor.\ Math.\ Phys.\  {\bf 2}, 231 (1998)
  [Int.\ J.\ Theor.\ Phys.\  {\bf 38}, 1113 (1999)]
  [ {\tt hep-th/9711200}];
  %%CITATION = IJTPB,38,1113;%%
  
 \bibitem{GKP}
    S.~S.~Gubser, I.~R.~Klebanov and A.~M.~Polyakov,
  %``Gauge theory correlators from non-critical string theory,''
  Phys.\ Lett.\  B {\bf 428}, 105 (1998)
  [arXiv:hep-th/9802109].
  %%CITATION = PHLTA,B428,105;%%
  
\bibitem{WittenAdS}
E.~Witten,
  %``Anti-de Sitter space and holography,''
  Adv.\ Theor.\ Math.\ Phys.\  {\bf 2}, 253 (1998)
  [ {\tt hep-th/9802150}].
  %%CITATION = 00203,2,253;%%
  
\bibitem{RS}
  L.~Randall and R.~Sundrum,
  %``A large mass hierarchy from a small extra dimension,''
  Phys.\ Rev.\ Lett.\  {\bf 83}, 3370 (1999)
  [arXiv:hep-ph/9905221].
  %%CITATION = PRLTA,83,3370;%%
  
 \bibitem{HoloPheno}
   N.~Arkani-Hamed, M.~Porrati and L.~Randall,
  %``Holography and phenomenology,''
  JHEP {\bf 0108}, 017 (2001)
  [arXiv:hep-th/0012148].
  %%CITATION = JHEPA,0108,017;%%
  
\bibitem{Csaki:2005vy}
  C.~Csaki, J.~Hubisz and P.~Meade,
  %``Electroweak symmetry breaking from extra dimensions,''
  arXiv:hep-ph/0510275.
  %%CITATION = HEP-PH/0510275;%%
  
\bibitem{KS}
  I.~R.~Klebanov and M.~J.~Strassler,
  %``Supergravity and a confining gauge theory: Duality cascades and
  %chiSB-resolution of naked singularities,''
  JHEP {\bf 0008}, 052 (2000)
  [arXiv:hep-th/0007191].
  %%CITATION = JHEPA,0008,052;%%
  
\bibitem{bulkgauge}
  H.~Davoudiasl, J.~L.~Hewett and T.~G.~Rizzo,
  %``Bulk gauge fields in the Randall-Sundrum model,''
  Phys.\ Lett.\  B {\bf 473}, 43 (2000)
  [arXiv:hep-ph/9911262];
  %%CITATION = PHLTA,B473,43;%%
   A.~Pomarol,
  %``Gauge bosons in a five-dimensional theory with localized gravity,''
  Phys.\ Lett.\  B {\bf 486}, 153 (2000)
  [arXiv:hep-ph/9911294];
  %%CITATION = PHLTA,B486,153;%%
  S.~Chang, J.~Hisano, H.~Nakano, N.~Okada and M.~Yamaguchi,
  %``Bulk standard model in the Randall-Sundrum background,''
  Phys.\ Rev.\  D {\bf 62}, 084025 (2000)
  [arXiv:hep-ph/9912498].
  %%CITATION = PHRVA,D62,084025;%%  
  
\bibitem{custodialRS}
  K.~Agashe, A.~Delgado, M.~J.~May and R.~Sundrum,
  %``RS1, custodial isospin and precision tests,''
  JHEP {\bf 0308}, 050 (2003)
  [arXiv:hep-ph/0308036].
  %%CITATION = JHEPA,0308,050;%%
  
 \bibitem{PomarolRunning}
   A.~Pomarol,
  %``Grand Unified Theories without the Desert,''
  Phys.\ Rev.\ Lett.\  {\bf 85}, 4004 (2000)
  [arXiv:hep-ph/0005293].
  %%CITATION = PRLTA,85,4004;%%
  
  %JHEP
\bibitem{Contino:2002kc}
  R.~Contino, P.~Creminelli and E.~Trincherini,
  %``Holographic evolution of gauge couplings,''
  JHEP {\bf 0210}, 029 (2002)
  [arXiv:hep-th/0208002].
  %%CITATION = JHEPA,0210,029;%% 
  
  %JHEP
\bibitem{Agashe:2005vg}
  K.~Agashe, R.~Contino and R.~Sundrum,
  %``Top compositeness and precision unification,''
  Phys.\ Rev.\ Lett.\  {\bf 95}, 171804 (2005)
  [arXiv:hep-ph/0502222].
  %%CITATION = PRLTA,95,171804;%%
  
  \bibitem{CreminelliEWPT}
   P.~Creminelli, A.~Nicolis and R.~Rattazzi,
  %``Holography and the electroweak phase transition,''
  JHEP {\bf 0203}, 051 (2002)
  [arXiv:hep-th/0107141].
  %%CITATION = JHEPA,0203,051;%%
  
  \bibitem{EmptyUniverse}
   J.~Kaplan, P.~C.~Schuster and N.~Toro,
  %``Avoiding an empty universe in RS I models and large-N gauge theories,''
  arXiv:hep-ph/0609012.
  %%CITATION = HEP-PH/0609012;%%
  
  \bibitem{Strassler:2003ht}
  M.~J.~Strassler,
  %``Non-supersymmetric theories with light scalar fields and large
  %hierarchies,''
  arXiv:hep-th/0309122.
  %%CITATION = HEP-TH/0309122;%%
  
\bibitem{Kachru:2009kg}
  S.~Kachru, D.~Simic and S.~P.~Trivedi,
  %``Stable Non-Supersymmetric Throats in String Theory,''
  arXiv:0905.2970 [hep-th].
  %%CITATION = ARXIV:0905.2970;%%
  
\bibitem{Strassler:2008bv}
  M.~J.~Strassler,
  %``Why Unparticle Models with Mass Gaps are Examples of Hidden Valleys,''
  arXiv:0801.0629 [hep-ph].
  %%CITATION = ARXIV:0801.0629;%%
  
\bibitem{Hassanain:2009at}
  B.~Hassanain, J.~March-Russell and J.~G.~Rosa,
  %``On the possibility of light string resonances at the LHC and Tevatron from
  %Randall-Sundrum throats,''
  JHEP {\bf 0907}, 077 (2009)
  [arXiv:0904.4108 [hep-ph]].
  %%CITATION = JHEPA,0907,077;%%
  
\bibitem{Perelstein:2009qi}
  M.~Perelstein and A.~Spray,
  %``Tensor Reggeons from Warped Space at the LHC,''
  JHEP {\bf 0910}, 096 (2009)
  [arXiv:0907.3496 [hep-ph]].
  %%CITATION = JHEPA,0910,096;%%
  
\bibitem{Deconstruction}
  N.~Arkani-Hamed, A.~G.~Cohen and H.~Georgi,
  %``(De)constructing dimensions,''
  Phys.\ Rev.\ Lett.\  {\bf 86}, 4757 (2001)
  [arXiv:hep-th/0104005].
  %%CITATION = PRLTA,86,4757;%%
  
\bibitem{DeconstructingRS}
  H.~C.~Cheng, C.~T.~Hill and J.~Wang,
  %``Dynamical electroweak breaking and latticized extra dimensions,''
  Phys.\ Rev.\  D {\bf 64}, 095003 (2001)
  [arXiv:hep-ph/0105323];
  %%CITATION = PHRVA,D64,095003;%%
  H.~Abe, T.~Kobayashi, N.~Maru and K.~Yoshioka,
  %``Field localization in warped gauge theories,''
  Phys.\ Rev.\  D {\bf 67}, 045019 (2003)
  [arXiv:hep-ph/0205344];
  %%CITATION = PHRVA,D67,045019;%%
    A.~Falkowski and H.~D.~Kim,
  %``Running of gauge couplings in AdS(5) via deconstruction,''
  JHEP {\bf 0208}, 052 (2002)
  [arXiv:hep-ph/0208058];
  %%CITATION = JHEPA,0208,052;%%
  L.~Randall, Y.~Shadmi and N.~Weiner,
  %``Deconstruction and Gauge Theories in AdS_5,''
  JHEP {\bf 0301}, 055 (2003)
  [arXiv:hep-th/0208120].
  %%CITATION = JHEPA,0301,055;%%
  
\bibitem{Seiberg:1994pq}
  N.~Seiberg,
  %``Electric - magnetic duality in supersymmetric nonAbelian gauge theories,''
  Nucl.\ Phys.\  B {\bf 435}, 129 (1995)
  [arXiv:hep-th/9411149].
  %%CITATION = NUPHA,B435,129;%%
    
 \bibitem{SMasComposite}
   M.~J.~Strassler,
  %``Duality in supersymmetric field theory: General conceptual background and
  %an application to real particle physics,''
%\href{http://www.slac.stanford.edu/spires/find/hep/www?irn=4969456}{SPIRES entry}
Prepared for International Workshop on Perspectives of Strong Coupling Gauge Theories (SCGT 96), Nagoya, Japan, 13-16 Nov 1996. Available at 
\verb!http://www.eken.phys.nagoya-u.ac.jp/Scgt/proc/!

\bibitem{SMcascade}
  J.~F.~G.~Cascales, F.~Saad and A.~M.~Uranga,
  %``Holographic dual of the standard model on the throat,''
  JHEP {\bf 0511}, 047 (2005)
  [arXiv:hep-th/0503079];
  %%CITATION = JHEPA,0511,047;%%
   J.~J.~Heckman, C.~Vafa, H.~Verlinde and M.~Wijnholt,
  %``Cascading to the MSSM,''
  JHEP {\bf 0806}, 016 (2008)
  [arXiv:0711.0387 [hep-ph]].
  %%CITATION = JHEPA,0806,016;%%
  
 \bibitem{LittleRS}
  H.~Davoudiasl, G.~Perez and A.~Soni,
  %``The Little Randall-Sundrum Model at the Large Hadron Collider,''
  Phys.\ Lett.\  B {\bf 665}, 67 (2008)
  [arXiv:0802.0203 [hep-ph]];
  %%CITATION = PHLTA,B665,67;%%
  H.~Davoudiasl, S.~Gopalakrishna and A.~Soni,
  %``Big Signals of Little Randall-Sundrum Models,''
  arXiv:0908.1131 [hep-ph].
  %%CITATION = ARXIV:0908.1131;%% 
  
\bibitem{GaiottoMaldacena}
  D.~Gaiotto and J.~Maldacena,
  %``The gravity duals of N=2 superconformal field theories,''
  arXiv:0904.4466 [hep-th].
  %%CITATION = ARXIV:0904.4466;%%
  
\bibitem{TReheat}
  S.~Hannestad,
  %``What is the lowest possible reheating temperature?,''
  Phys.\ Rev.\  D {\bf 70}, 043506 (2004)
  [arXiv:astro-ph/0403291].
  %%CITATION = PHRVA,D70,043506;%%
  
\bibitem{HawkingPage}
  S.~W.~Hawking and D.~N.~Page,
  %``Thermodynamics Of Black Holes In Anti-De Sitter Space,''
  Commun.\ Math.\ Phys.\  {\bf 87}, 577 (1983).
  %%CITATION = CMPHA,87,577;%%
  
 \bibitem{WittenBH}
  E.~Witten,
  %``Anti-de Sitter space, thermal phase transition, and confinement in  gauge
  %theories,''
  Adv.\ Theor.\ Math.\ Phys.\  {\bf 2}, 505 (1998)
  [arXiv:hep-th/9803131].
  %%CITATION = 00203,2,505;%%
  
\bibitem{Herzog:2006ra}
  C.~P.~Herzog,
  %``A holographic prediction of the deconfinement temperature,''
  Phys.\ Rev.\ Lett.\  {\bf 98}, 091601 (2007)
  [arXiv:hep-th/0608151].
  %%CITATION = PRLTA,98,091601;%%
  
\bibitem{BallonBayona:2007vp}
  C.~A.~Ballon Bayona, H.~Boschi-Filho, N.~R.~F.~Braga and L.~A.~Pando Zayas,
  %``On a holographic model for confinement / deconfinement,''
  Phys.\ Rev.\  D {\bf 77}, 046002 (2008)
  [arXiv:0705.1529 [hep-th]].
  %%CITATION = PHRVA,D77,046002;%%
 
 \bibitem{RandallServant}
  L.~Randall and G.~Servant,
  %``Gravitational Waves from Warped Spacetime,''
  JHEP {\bf 0705}, 054 (2007)
  [arXiv:hep-ph/0607158].
  %%CITATION = JHEPA,0705,054;%%
  
\bibitem{Nardini:2007me}
  G.~Nardini, M.~Quiros and A.~Wulzer,
  %``A Confining Strong First-Order Electroweak Phase Transition,''
  JHEP {\bf 0709}, 077 (2007)
  [arXiv:0706.3388 [hep-ph]].
  %%CITATION = JHEPA,0709,077;%%
  
 \bibitem{Hassanain:2007js}
  B.~Hassanain, J.~March-Russell and M.~Schvellinger,
  %``Warped Deformed Throats have Faster (Electroweak) Phase Transitions,''
  JHEP {\bf 0710}, 089 (2007)
  [arXiv:0708.2060 [hep-th]].
  %%CITATION = JHEPA,0710,089;%%
  
\bibitem{Henningson:1998gx}
  M.~Henningson and K.~Skenderis,
  %``The holographic Weyl anomaly,''
  JHEP {\bf 9807}, 023 (1998)
  [arXiv:hep-th/9806087].
  %%CITATION = JHEPA,9807,023;%%
  
 \bibitem{KarchKatz}
  A.~Karch and E.~Katz,
  %``Adding flavor to AdS/CFT,''
  JHEP {\bf 0206}, 043 (2002)
  [arXiv:hep-th/0205236].
  %%CITATION = JHEPA,0206,043;%%

%JHEP  
\bibitem{Gherghetta:2006yq}
  T.~Gherghetta and J.~Giedt,
  %``Bulk fields in AdS(5) from probe D7 branes,''
  Phys.\ Rev.\  D {\bf 74}, 066007 (2006)
  [arXiv:hep-th/0605212].
  %%CITATION = PHRVA,D74,066007;%%
  
   \bibitem{Gubser:1998vd}
  S.~S.~Gubser,
  %``Einstein manifolds and conformal field theories,''
  Phys.\ Rev.\  D {\bf 59}, 025006 (1999)
  [arXiv:hep-th/9807164].
  %%CITATION = PHRVA,D59,025006;%%
  
 \bibitem{PS1}
  %%CITATION = JHEPA,0002,006;%%
  J.~Polchinski and M.~J.~Strassler,
  %``The string dual of a confining four-dimensional gauge theory,''
  arXiv:hep-th/0003136.
  %%CITATION = HEP-TH/0003136;%%
  
\bibitem{DualityCascadeLectures}
  M.~J.~Strassler,
  %``The duality cascade,''
  arXiv:hep-th/0505153.
  %%CITATION = HEP-TH/0505153;%%
  
\bibitem{D7FlavorInThroat}
  F.~Benini, F.~Canoura, S.~Cremonesi, C.~Nunez and A.~V.~Ramallo,
  %``Backreacting Flavors in the Klebanov-Strassler Background,''
  JHEP {\bf 0709}, 109 (2007)
  [arXiv:0706.1238 [hep-th]];
  %%CITATION = JHEPA,0709,109;%%
  F.~Bigazzi, A.~L.~Cotrone, A.~Paredes and A.~V.~Ramallo,
  %``The Klebanov-Strassler model with massive dynamical flavors,''
  JHEP {\bf 0903}, 153 (2009)
  [arXiv:0812.3399 [hep-th]].
  %%CITATION = JHEPA,0903,153;%%
  
\bibitem{orbifold}
  S.~Kachru and E.~Silverstein,
  %``4d conformal theories and strings on orbifolds,''
  Phys.\ Rev.\ Lett.\  {\bf 80}, 4855 (1998)
  [arXiv:hep-th/9802183];
  %%CITATION = PRLTA,80,4855;%%
    A.~E.~Lawrence, N.~Nekrasov and C.~Vafa,
  %``On conformal field theories in four dimensions,''
  Nucl.\ Phys.\  B {\bf 533}, 199 (1998)
  [arXiv:hep-th/9803015];
  %%CITATION = NUPHA,B533,199;%%
  M.~Bershadsky, Z.~Kakushadze and C.~Vafa,
  %``String expansion as large N expansion of gauge theories,''
  Nucl.\ Phys.\  B {\bf 523}, 59 (1998)
  [arXiv:hep-th/9803076];
  %%CITATION = NUPHA,B523,59;%%
   Y.~Oz and J.~Terning,
  %``Orbifolds of AdS(5) x S(5) and 4d conformal field theories,''
  Nucl.\ Phys.\  B {\bf 532}, 163 (1998)
  [arXiv:hep-th/9803167].
  %%CITATION = NUPHA,B532,163;%%
  M.~Bershadsky and A.~Johansen,
  %``Large N limit of orbifold field theories,''
  Nucl.\ Phys.\  B {\bf 536}, 141 (1998)
  [arXiv:hep-th/9803249].
  %%CITATION = NUPHA,B536,141;%%
  
\bibitem{HOP}
  G.~T.~Horowitz, J.~Orgera and J.~Polchinski,
  %``Nonperturbative Instability of AdS_5 x S^5/Z_k,''
  Phys.\ Rev.\  D {\bf 77}, 024004 (2008)
  [arXiv:0709.4262 [hep-th]].
  %%CITATION = PHRVA,D77,024004;%%
  
\bibitem{Witten:1981gj}
  E.~Witten,
  %``Instability Of The Kaluza-Klein Vacuum,''
  Nucl.\ Phys.\  B {\bf 195}, 481 (1982).
  %%CITATION = NUPHA,B195,481;%%
  
\bibitem{Ypq}
  J.~P.~Gauntlett, D.~Martelli, J.~Sparks and D.~Waldram,
  %``Sasaki-Einstein metrics on S(2) x S(3),''
  Adv.\ Theor.\ Math.\ Phys.\  {\bf 8}, 711 (2004)
  [arXiv:hep-th/0403002];
  %%CITATION = 00203,8,711;%%
  D.~Martelli and J.~Sparks,
  %``Toric geometry, Sasaki-Einstein manifolds and a new infinite class of
  %AdS/CFT duals,''
  Commun.\ Math.\ Phys.\  {\bf 262}, 51 (2006)
  [arXiv:hep-th/0411238].
  %%CITATION = CMPHA,262,51;%%
  
\bibitem{Herzog:2004tr}
  C.~P.~Herzog, Q.~J.~Ejaz and I.~R.~Klebanov,
  %``Cascading RG flows from new Sasaki-Einstein manifolds,''
  JHEP {\bf 0502}, 009 (2005)
  [arXiv:hep-th/0412193].
  %%CITATION = JHEPA,0502,009;%%

\bibitem{PolchinskiSilverstein}
  J.~Polchinski and E.~Silverstein,
  %``Dual Purpose Landscaping Tools: Small Extra Dimensions in AdS/CFT,''
  arXiv:0908.0756 [hep-th].
  %%CITATION = ARXIV:0908.0756;%%
  
\bibitem{SakaiSugimoto}
  T.~Sakai and S.~Sugimoto,
  %``Low energy hadron physics in holographic QCD,''
  Prog.\ Theor.\ Phys.\  {\bf 113}, 843 (2005)
  [arXiv:hep-th/0412141].
  %%CITATION = PTPKA,113,843;%%
  
\bibitem{SakaiSugimoto2}
  T.~Sakai and S.~Sugimoto,
  %``More on a holographic dual of QCD,''
  Prog.\ Theor.\ Phys.\  {\bf 114}, 1083 (2005)
  [arXiv:hep-th/0507073].
  %%CITATION = PTPKA,114,1083;%%
  
\bibitem{Carone:2007md}
  C.~D.~Carone, J.~Erlich and M.~Sher,
  %``Holographic Electroweak Symmetry Breaking from D-branes,''
  Phys.\ Rev.\  D {\bf 76}, 015015 (2007)
  [arXiv:0704.3084 [hep-th]].
  %%CITATION = PHRVA,D76,015015;%%
  
 \bibitem{Hirayama:2007hz}
  T.~Hirayama and K.~Yoshioka,
  %``Holographic Construction of Technicolor Theory,''
  JHEP {\bf 0710}, 002 (2007)
  [arXiv:0705.3533 [hep-ph]].
  %%CITATION = JHEPA,0710,002;%%
  
\bibitem{Carone:2008rx}
  C.~D.~Carone, J.~Erlich and M.~Sher,
  %``Extra Gauge Invariance from an Extra Dimension,''
  Phys.\ Rev.\  D {\bf 78}, 015001 (2008)
  [arXiv:0802.3702 [hep-ph]].
  %%CITATION = PHRVA,D78,015001;%%
  
\bibitem{Mintakevich:2009wz}
  O.~Mintakevich and J.~Sonnenschein,
  %``Holographic technicolor models and their S-parameter,''
  JHEP {\bf 0907}, 032 (2009)
  [arXiv:0905.3284 [hep-th]].
  %%CITATION = JHEPA,0907,032;%%
  
\bibitem{Antonyan:2006vw}
  E.~Antonyan, J.~A.~Harvey, S.~Jensen and D.~Kutasov,
  %``NJL and QCD from string theory,''
  arXiv:hep-th/0604017.
  %%CITATION = HEP-TH/0604017;%%
  
 \bibitem{HiddenValley}
  M.~J.~Strassler and K.~M.~Zurek,
  %``Echoes of a hidden valley at hadron colliders,''
  Phys.\ Lett.\  B {\bf 651}, 374 (2007)
  [arXiv:hep-ph/0604261].
  %%CITATION = PHLTA,B651,374;%%
  
\bibitem{Quirks}
  L.~B.~Okun,
  %``Thetons,''
  JETP Lett.\  {\bf 31}, 144 (1980)
  [Pisma Zh.\ Eksp.\ Teor.\ Fiz.\  {\bf 31}, 156 (1979)];
  %%CITATION = ZFPRA,31,156;%%
    L.~B.~Okun,
  %``Theta Particles,''
  Nucl.\ Phys.\  B {\bf 173}, 1 (1980);
  %%CITATION = NUPHA,B173,1;%%
  J.~Kang and M.~A.~Luty,
  %``Macroscopic Strings and 'Quirks' at Colliders,''
  JHEP {\bf 0911}, 065 (2009)
  [arXiv:0805.4642 [hep-ph]].
  %%CITATION = JHEPA,0911,065;%%
  
\bibitem{GlueHiddenValley}
  J.~E.~Juknevich, D.~Melnikov and M.~J.~Strassler,
  %``A Pure-Glue Hidden Valley I. States and Decays,''
  JHEP {\bf 0907}, 055 (2009)
  [arXiv:0903.0883 [hep-ph]].
  %%CITATION = JHEPA,0907,055;%%
  
\bibitem{SakaiSugimotoFiniteT}
  O.~Aharony, J.~Sonnenschein and S.~Yankielowicz,
  %``A holographic model of deconfinement and chiral symmetry restoration,''
  Annals Phys.\  {\bf 322}, 1420 (2007)
  [arXiv:hep-th/0604161];
  %%CITATION = APNYA,322,1420;%%
  A.~Parnachev and D.~A.~Sahakyan,
  %``Chiral phase transition from string theory,''
  Phys.\ Rev.\ Lett.\  {\bf 97}, 111601 (2006)
  [arXiv:hep-th/0604173].
  %%CITATION = PRLTA,97,111601;%%
  
\bibitem{Gaiotto:2009we}
  D.~Gaiotto,
  %``N=2 dualities,''
  arXiv:0904.2715 [hep-th].
  %%CITATION = ARXIV:0904.2715;%%
  
\bibitem{weakgravity}
  N.~Arkani-Hamed, L.~Motl, A.~Nicolis and C.~Vafa,
  %``The string landscape, black holes and gravity as the weakest force,''
  JHEP {\bf 0706}, 060 (2007)
  [arXiv:hep-th/0601001].
  %%CITATION = JHEPA,0706,060;%%

 \bibitem{KlebanovMaldacena}
   I.~R.~Klebanov and J.~M.~Maldacena,
  %``Superconformal gauge theories and non-critical superstrings,''
  Int.\ J.\ Mod.\ Phys.\  A {\bf 19}, 5003 (2004)
  [arXiv:hep-th/0409133].
  %%CITATION = IMPAE,A19,5003;%%

 \bibitem{confinement}
     G.~'t Hooft,
  %``A PLANAR DIAGRAM THEORY FOR STRONG INTERACTIONS,''
  Nucl.\ Phys.\  B {\bf 72}, 461 (1974);
  %%CITATION = NUPHA,B72,461;%%
   K.~G.~Wilson,
  %``CONFINEMENT OF QUARKS,''
  Phys.\ Rev.\  D {\bf 10}, 2445 (1974);
  %%CITATION = PHRVA,D10,2445;%%
    Y.~Nambu,
  %``Strings, monopoles, and gauge fields,''
  Phys.\ Rev.\  D {\bf 10}, 4262 (1974);
  %%CITATION = PHRVA,D10,4262;%%
   G.~'t Hooft,
  %``On The Phase Transition Towards Permanent Quark Confinement,''
  Nucl.\ Phys.\  B {\bf 138}, 1 (1978).
  %%CITATION = NUPHA,B138,1;%%
   Y.~Nambu,
  %``QCD And The String Model,''
  Phys.\ Lett.\  B {\bf 80}, 372 (1979);
  %%CITATION = PHLTA,B80,372;%%
    M.~Luscher, G.~Munster and P.~Weisz,
  %``How Thick Are Chromoelectric Flux Tubes?,''
  Nucl.\ Phys.\  B {\bf 180}, 1 (1981);
  %%CITATION = NUPHA,B180,1;%%
   M.~Luscher,
  %``Symmetry Breaking Aspects Of The Roughening Transition In Gauge Theories,''
  Nucl.\ Phys.\  B {\bf 180}, 317 (1981);
  %%CITATION = NUPHA,B180,317;%%
  R.~Sundrum,
  %``Hadronic string from confinement,''
  arXiv:hep-ph/9702306.
  %%CITATION = HEP-PH/9702306;%%

\bibitem{resum}
  J.~K.~Erickson, G.~W.~Semenoff, R.~J.~Szabo and K.~Zarembo,
  %``Static potential in N = 4 supersymmetric Yang-Mills theory,''
  Phys.\ Rev.\  D {\bf 61}, 105006 (2000)
  [arXiv:hep-th/9911088];
  %%CITATION = PHRVA,D61,105006;%%
    J.~K.~Erickson, G.~W.~Semenoff and K.~Zarembo,
  %``Wilson loops in N = 4 supersymmetric Yang-Mills theory,''
  Nucl.\ Phys.\  B {\bf 582}, 155 (2000)
  [arXiv:hep-th/0003055]
  %%CITATION = NUPHA,B582,155;%%
  
\bibitem{Bachas}
  C.~Bachas,
  %``Convexity Of The Quarkonium Potential,''
  Phys.\ Rev.\  D {\bf 33}, 2723 (1986).
  %%CITATION = PHRVA,D33,2723;%%  

\bibitem{techniEW}  
M.~E.~Peskin and T.~Takeuchi,
%``A New constraint on a strongly interacting Higgs sector,''
Phys.\ Rev.\ Lett.\  {\bf 65}, 964 (1990);
%%CITATION = PRLTA,65,964;%%
B.~Holdom and J.~Terning,
%``Large corrections to electroweak parameters in technicolor theories,''
Phys.\ Lett.\  B {\bf 247}, 88 (1990);
%%CITATION = PHLTA,B247,88;%%
  M.~Golden and L.~Randall,
%``RADIATIVE CORRECTIONS TO ELECTROWEAK PARAMETERS IN TECHNICOLOR THEORIES,''
Nucl.\ Phys.\  B {\bf 361}, 3 (1991);
%%CITATION = NUPHA,B361,3;%%
  M.~E.~Peskin and T.~Takeuchi,
%``Estimation of oblique electroweak corrections,''
Phys.\ Rev.\  D {\bf 46}, 381 (1992).  %%CITATION = PHRVA,D46,381;%%
  
\bibitem{CsakiErlichTerning}
  C.~Csaki, J.~Erlich and J.~Terning,
  %``The effective Lagrangian in the Randall-Sundrum model and electroweak
  %physics,''
  Phys.\ Rev.\  D {\bf 66}, 064021 (2002)
  [arXiv:hep-ph/0203034].
  %%CITATION = PHRVA,D66,064021;%%
  
\bibitem{Disguising}
  C.~Grojean, W.~Skiba and J.~Terning,
  %``Disguising the oblique parameters,''
  Phys.\ Rev.\  D {\bf 73}, 075008 (2006)
  [arXiv:hep-ph/0602154].
  %%CITATION = PHRVA,D73,075008;%%
  
\bibitem{Cacciapaglia:2006pk}
  G.~Cacciapaglia, C.~Csaki, G.~Marandella and A.~Strumia,
  %``The minimal set of electroweak precision parameters,''
  Phys.\ Rev.\  D {\bf 74}, 033011 (2006)
  [arXiv:hep-ph/0604111].
  %%CITATION = PHRVA,D74,033011;%%
  
\bibitem{HiggslessOblique}
    G.~Cacciapaglia, C.~Csaki, C.~Grojean and J.~Terning,
  %``Oblique corrections from Higgsless models in warped space,''
  Phys.\ Rev.\  D {\bf 70}, 075014 (2004)
  [arXiv:hep-ph/0401160]
  %%CITATION = PHRVA,D70,075014;%%
  
\bibitem{positiveS}
  R.~Barbieri, A.~Pomarol and R.~Rattazzi,
  %``Weakly coupled Higgsless theories and precision electroweak tests,''
  Phys.\ Lett.\  B {\bf 591}, 141 (2004)
  [arXiv:hep-ph/0310285];
  %%CITATION = PHLTA,B591,141;%%
  D.~K.~Hong and H.~U.~Yee,
  %``Holographic estimate of oblique corrections for technicolor,''
  Phys.\ Rev.\  D {\bf 74}, 015011 (2006)
  [arXiv:hep-ph/0602177];
  %%CITATION = PHRVA,D74,015011;%%
  K.~Agashe, C.~Csaki, C.~Grojean and M.~Reece,
  %``The S-parameter in holographic technicolor models,''
  JHEP {\bf 0712}, 003 (2007)
  [arXiv:0704.1821 [hep-ph]].
  %%CITATION = JHEPA,0712,003;%%  

\bibitem{delocalized}
  G.~Cacciapaglia, C.~Csaki, C.~Grojean and J.~Terning,
  %``Curing the Ills of Higgsless models: The S parameter and unitarity,''
  Phys.\ Rev.\  D {\bf 71}, 035015 (2005)
  [arXiv:hep-ph/0409126];
  %%CITATION = PHRVA,D71,035015;%%
  R.~Foadi, S.~Gopalakrishna and C.~Schmidt,
  %``Effects of fermion localization in Higgsless theories and electroweak
  %constraints,''
  Phys.\ Lett.\  B {\bf 606}, 157 (2005)
  [arXiv:hep-ph/0409266];
  %%CITATION = PHLTA,B606,157;%%
  R.~S.~Chivukula, E.~H.~Simmons, H.~J.~He, M.~Kurachi and M.~Tanabashi,
  %``Deconstructed Higgsless models with one-site delocalization,''
  Phys.\ Rev.\  D {\bf 71}, 115001 (2005)
  [arXiv:hep-ph/0502162].
  %%CITATION = PHRVA,D71,115001;%%
  
\bibitem{Kaplan:1991dc}
  D.~B.~Kaplan,
  %``Flavor at SSC energies: A New mechanism for dynamically generated fermion
  %masses,''
  Nucl.\ Phys.\  B {\bf 365}, 259 (1991).
  %%CITATION = NUPHA,B365,259;%%
   
\bibitem{AdSFermions}
  R.~Contino and A.~Pomarol,
  %``Holography for fermions,''
  JHEP {\bf 0411}, 058 (2004)
  [arXiv:hep-th/0406257].
  %%CITATION = JHEPA,0411,058;%%

%JHEP -- added Gherghetta 
\bibitem{RSflavor}
  Y.~Grossman and M.~Neubert,
  %``Neutrino masses and mixings in non-factorizable geometry,''
  Phys.\ Lett.\  B {\bf 474}, 361 (2000)
  [arXiv:hep-ph/9912408];
  %%CITATION = PHLTA,B474,361;%%
  T.~Gherghetta and A.~Pomarol,
  %``Bulk fields and supersymmetry in a slice of AdS,''
  Nucl.\ Phys.\  B {\bf 586}, 141 (2000)
  [arXiv:hep-ph/0003129];
  %%CITATION = NUPHA,B586,141;%%
  K.~Agashe, G.~Perez and A.~Soni,
  %``Flavor structure of warped extra dimension models,''
  Phys.\ Rev.\  D {\bf 71}, 016002 (2005)
  [arXiv:hep-ph/0408134];
  %%CITATION = PHRVA,D71,016002;%%
  K.~Agashe, M.~Papucci, G.~Perez and D.~Pirjol,
  %``Next to minimal flavor violation,''
  arXiv:hep-ph/0509117;
  %%CITATION = HEP-PH/0509117;%%
   G.~Cacciapaglia, C.~Csaki, J.~Galloway, G.~Marandella, J.~Terning and A.~Weiler,
  %``A GIM Mechanism from Extra Dimensions,''
  JHEP {\bf 0804}, 006 (2008)
  [arXiv:0709.1714 [hep-ph]];
  %%CITATION = JHEPA,0804,006;%%
  A.~L.~Fitzpatrick, G.~Perez and L.~Randall,
  %``Flavor from Minimal Flavor Violation & a Viable Randall-Sundrum Model,''
  arXiv:0710.1869 [hep-ph];
  %%CITATION = ARXIV:0710.1869;%%
  C.~Csaki, A.~Falkowski and A.~Weiler,
  %``The Flavor of the Composite Pseudo-Goldstone Higgs,''
  JHEP {\bf 0809}, 008 (2008)
  [arXiv:0804.1954 [hep-ph]];
  %%CITATION = JHEPA,0809,008;%%
  J.~Santiago,
  %``Minimal Flavor Protection: A New Flavor Paradigm in Warped Models,''
  JHEP {\bf 0812}, 046 (2008)
  [arXiv:0806.1230 [hep-ph]].
  %%CITATION = JHEPA,0812,046;%%
  M.~Blanke, A.~J.~Buras, B.~Duling, S.~Gori and A.~Weiler,
  %``$\Delta$ F=2 Observables and Fine-Tuning in a Warped Extra Dimension with
  %Custodial Protection,''
  JHEP {\bf 0903}, 001 (2009)
  [arXiv:0809.1073 [hep-ph]];
  %%CITATION = JHEPA,0903,001;%%
    K.~Agashe, A.~Azatov and L.~Zhu,
  %``Flavor Violation Tests of Warped/Composite SM in the Two-Site Approach,''
  Phys.\ Rev.\  D {\bf 79}, 056006 (2009)
  [arXiv:0810.1016 [hep-ph]];
  %%CITATION = PHRVA,D79,056006;%%
  C.~Csaki, G.~Perez, Z.~Surujon and A.~Weiler,
  %``Flavor Alignment via Shining in RS,''
  arXiv:0907.0474 [hep-ph].
  %%CITATION = ARXIV:0907.0474;%%
  
\bibitem{string_collider}
  S.~Cullen, M.~Perelstein and M.~E.~Peskin,
  %``TeV strings and collider probes of large extra dimensions,''
  Phys.\ Rev.\  D {\bf 62}, 055012 (2000)
  [arXiv:hep-ph/0001166];
  %%CITATION = PHRVA,D62,055012;%%
  I.~Antoniadis, K.~Benakli and A.~Laugier,
  %``Contact interactions in D-brane models,''
  JHEP {\bf 0105}, 044 (2001)
  [arXiv:hep-th/0011281];
  %%CITATION = JHEPA,0105,044;%%
  S.~Dimopoulos and R.~Emparan,
  %``String balls at the LHC and beyond,''
  Phys.\ Lett.\  B {\bf 526}, 393 (2002)
  [arXiv:hep-ph/0108060];
  %%CITATION = PHLTA,B526,393;%%
 P.~Burikham, T.~Han, F.~Hussain and D.~W.~McKay,
  %``Bounds on four fermion contact interactions induced by string resonances,''
  Phys.\ Rev.\  D {\bf 69}, 095001 (2004)
  [arXiv:hep-ph/0309132];
  %%CITATION = PHRVA,D69,095001;%%
  P.~Burikham, T.~Figy and T.~Han,
  %``TeV-scale string resonances at hadron colliders,''
  Phys.\ Rev.\  D {\bf 71}, 016005 (2005)
  [Erratum-ibid.\  D {\bf 71}, 019905 (2005)]
  [arXiv:hep-ph/0411094];
  %%CITATION = PHRVA,D71,016005;%%
  P.~Meade and L.~Randall,
  %``Black Holes and Quantum Gravity at the LHC,''
  JHEP {\bf 0805}, 003 (2008)
  [arXiv:0708.3017 [hep-ph]];
  %%CITATION = JHEPA,0805,003;%%
   For a recent review, see   D.~Lust,
  %``Seeing through the String Landscape - a String Hunter's Companion in
  %Particle Physics and Cosmology,''
  JHEP {\bf 0903}, 149 (2009)
  [arXiv:0904.4601 [hep-th]].
  %%CITATION = JHEPA,0903,149;%%

\bibitem{glueball}
A.~L.~Fitzpatrick, J.~Kaplan, L.~Randall and L.~T.~Wang,
  %``Searching for the Kaluza-Klein Graviton in Bulk RS Models,''
  JHEP {\bf 0709}, 013 (2007)
  [arXiv:hep-ph/0701150].
  %%CITATION = JHEPA,0709,013;%%
  K.~Agashe, H.~Davoudiasl, G.~Perez and A.~Soni,
  %``Warped Gravitons at the LHC and Beyond,''
  Phys.\ Rev.\  D {\bf 76}, 036006 (2007)
  [arXiv:hep-ph/0701186].
  %%CITATION = PHRVA,D76,036006;%%

\bibitem{PolchinskiStrasslerDIS}
  J.~Polchinski and M.~J.~Strassler,
  %``Deep inelastic scattering and gauge/string duality,''
  JHEP {\bf 0305}, 012 (2003)
  [arXiv:hep-th/0209211].
  %%CITATION = JHEPA,0305,012;%%

 \bibitem{PolchinskiStrassler}
   J.~Polchinski and M.~J.~Strassler,
  %``Hard scattering and gauge/string duality,''
  Phys.\ Rev.\ Lett.\  {\bf 88}, 031601 (2002)
  [arXiv:hep-th/0109174];
  %%CITATION = PRLTA,88,031601;%%
 R.~C.~Brower, J.~Polchinski, M.~J.~Strassler and C.~I.~Tan,
  %``The Pomeron and Gauge/String Duality,''
  JHEP {\bf 0712}, 005 (2007)
  [arXiv:hep-th/0603115];
  %%CITATION = JHEPA,0712,005;%%
  R.~C.~Brower, M.~J.~Strassler and C.~I.~Tan,
  %``On The Pomeron at Large 't Hooft Coupling,''
  arXiv:0710.4378 [hep-th].
  %%CITATION = ARXIV:0710.4378;%%
  
\bibitem{ads_qcd}
D.~M.~Hofman and J.~Maldacena,
  %``Conformal collider physics: Energy and charge correlations,''
  JHEP {\bf 0805}, 012 (2008)
  [arXiv:0803.1467 [hep-th]].
  %%CITATION = JHEPA,0805,012;%%
  Y.~Hatta, E.~Iancu and A.~H.~Mueller,
  %``Jet evolution in the N=4 SYM plasma at strong coupling,''
  JHEP {\bf 0805}, 037 (2008)
  [arXiv:0803.2481 [hep-th]].
  %%CITATION = JHEPA,0805,037;%%
  C.~Csaki, M.~Reece and J.~Terning,
  %``The AdS/QCD Correspondence: Still Undelivered,''
  JHEP {\bf 0905}, 067 (2009)
  [arXiv:0811.3001 [hep-ph]].
  %%CITATION = JHEPA,0905,067;%%
    
\bibitem{WatsonBessel}
  G.~N.~Watson, 
  {\it A Treatise on the Theory of Bessel Functions}, 
  Cambridge, UK: Univ. Pr. (1922) 816p.
  
\bibitem{Pak:2009em}
  M.~Pak and H.~Reinhardt,
  %``The Wilson loop from a Dyson equation,''
  Phys.\ Rev.\  D {\bf 80}, 125022 (2009)
  [arXiv:0910.2916 [hep-th]].
  %%CITATION = PHRVA,D80,125022;%%
  
\bibitem{DrukkerGross}
    N.~Drukker and D.~J.~Gross,
  %``An exact prediction of N = 4 SUSYM theory for string theory,''
  J.\ Math.\ Phys.\  {\bf 42}, 2896 (2001)
  [arXiv:hep-th/0010274].
  %%CITATION = JMAPA,42,2896;%%
  
\end{thebibliography}
\end{document}